\documentclass[11pt,a4paper]{article}

\usepackage{geometry}
	\usepackage{graphicx}
	\usepackage{blindtext}
	\usepackage{amsmath}
	\usepackage{graphicx}
	\usepackage{pgf,pgfarrows,pgfnodes,pgfautomata,pgfheaps,pgfshade, bbding}
	\usepackage{amsmath, amssymb, textcomp, txfonts, amsfonts}
	\usepackage{colortbl}
	\usepackage{color} 
	\usepackage[english]{babel}
	\usepackage{enumerate}
	\usepackage{url}
	\usepackage{hyperref}
	\usepackage{calrsfs}
	\usepackage{appendix}
	\usepackage{tikz}
	\usepackage{color}
	
	\usetikzlibrary{arrows}

	\newtheorem{dfn}{Definition}[section]
	
	\newtheorem{proposition}{Proposition}[section]
	\newtheorem{lem}{Lemma}[section]
	\newtheorem{thm}{Theorem}[section]
	\newtheorem{cor}{Corollary}[thm]
	\newtheorem{rem}{Remark}[section]
	\newtheorem{example}{Example}[section]

	\def\F{\mathbb F}

	\def\mc{\mathcal}

	\newcommand{\wh}{\color {white}}

\begin{document}

	\begin{titlepage}
	
	{\wh$\cdot$}
	
	\vspace{10cm}
	
		\begin{center}
			{\Huge Code based Cryptography: \\Classic McEliece}\\
			\vspace{2cm}
			{\Large Harshdeep Singh}\\
			{Scientific Analysis Group\\Defence R\&D Organisation, Delhi -- 110 054.
				\\{\wh .}\\\textcolor[rgb]{0,0,1}{harshdeep@sag.drdo.in}}
			\date{}
		\end{center}
		
	\end{titlepage}
	\pagebreak
	
	\tableofcontents
%
%
%
%

\subsection*{Introduction}
\addcontentsline{toc}{section}{Introduction}

{\Large T}he historical origins of coding theory are in the problem of reliable communication over noisy channels. Claude Shannon, in the introduction to his classic paper, ``A Mathematical Theory of Communication'' wrote \textit{The fundamental problem of communication is that of reproducing at one point either exactly or approximately a message selected at another point} in 1948. Error-correcting codes are widely used in applications such as returning pictures from deep space, design of numbers on Debit/Credit cards, ISBN on books, telephone numbers generation, cryptography and many more. It is fascinating to note the crucial role played by mathematics in successful deployment of those.

The progress of cryptography is closely related with the development of coding theory. In late 1970s, analogous to RSA, coding theory also started shaping public key cryptography.
Robert J. McEliece, in 1978, introduced a public-key cryptosystem based upon encoding the plaintext as codewords of an error correcting code from the family of Goppa codes\cite{rjm}.
In the originally proposed system, a codeword is generated from plaintext message bits by using a permuted and scrambled generator matrix of a Goppa code of length $n$, capable of correcting $t$ errors. This matrix is the public key. 
In this system, the ciphertext is formed by adding a randomly chosen error vector, containing some fixed number of non-zero bits, to each codeword of perturbed code.
The unperturbed Goppa code, together with scrambler and permutation matrices form the private key.
On reception, the associated private key is used to invoke an error-correcting decoder based upon the underlying Goppa code to correct the garbled bits in the codeword.

Due to inadequate and less effectiveness, code-based cryptosystems never came for widespread; though it is now being studied that these have not been directly affected by the algorithms developed by Peter Shor and Lov Grover. 
Some of these cryptosystems are proved to have substantial cryptographic strength which can be modified to achieve high security standards.
This provides ample margin against advances in super computing, including quantum computers.

Based on error-correcting codes and the difficult problem of decoding a message with random errors, the security of McEliece cryptosystem does not depend on the difficulty of factoring integers or finding the discrete logarithm of a number like in RSA, ElGamal and other well known cryptosystems. 
The security of these well known public key cryptosystems is at risk once quantum computers come to effect.
These computers not only promise to provide an enormous leap in computing power available to attackers but they effectively attack the heart of these well known cryptosystems, i.e. the problem to factor large integers and to solve discrete logarithms.

The security in McEliece cryptosystem lies on the ability of recovering plaintexts from ciphertexts, using a hidden error-correcting code, which the sender initially garbles with random errors. 
Quantum computers do not seem to give any significant improvements in attacking code-based systems, beyond the improvement in brute force search possible with Grover's algorithm. Therefore McEliece encryption scheme is one of the interesting code-based candidates for post-quantum cryptography. 
Due to problem of its large key sizes, this encryption scheme is modified a number of times. 
One such modification includes a ``dual'' variant of Generalized Reed Solomon codes, namely, Niederreiter in 1986. It improved the key size issue \cite{nie} which gave the speedups in software \cite{DJBSoft} and hardware implementations\cite{wan}. 
In 2017, Daniel J. Bernstein et al. proposed Classic McEliece, which is a code based post-quantum public key cryptosystem (PKC) candidate for NIST's global standardization.

The security level of McEliece cryptosystem has persisted outstandingly stable, despite a lot of attack papers over $40$ years. This resulted in improving efficiency and extending one-way chosen plaintext attacks (OW-CPA) to indistinguishability against adaptive chosen ciphertexts attack (IND-CCA2) security.

This report is primarily projected in such a way that the reader remains connected to main topic while covering necessary basics and fundamentals.
Chapter $1$ covers preliminaries and mathematical background required for understanding the primary objective of this report. It includes theory of finite fields, rings of polynomials, coding theory, error-correcting codes like Goppa codes including their encoding and decoding.
Then following Chapter $2$ begins by introducing coding theory based cryptosystems, which also includes the hard problems that roots the code based cryptography. We give a brief overview of information-set decoding (ISD) attack which can be applied on majority of code based cryptosystems.
Chapter $3$ covers the original McEliece cryptosystem based upon binary Goppa codes, with some attacks which can be applied on this scheme. Alongside, we describe the key size of the parameters imparted in the literature.
Chapter $4$ explains dual variant of McEliece cryptosystem, viz. Niederreiter scheme.
Then we arrive at key encapsulation mechanism, namely, Classic McEliece in chapter $5$ and cover the attacks and weaknesses of this code-based cryptosystem.
The following chapter $6$ indicates the strength of this cryptosystem addressing the points which lead to IND-CCA2 security.
Then we conclude this report in chapter $7$ describing some possibilities in future work. We have successfully implemented ISD attacks on small parameters set on McEliece cryptosystem in Appendix.

\section{Preliminaries}

The birth of coding theory was inspired by work of Golay, Hamming and Shannon in late $1940$'s. 
Coding theory is a field of study concerned with the transmission of data across noisy channels and the recovery of corrupted messages. 
Equivalently, coding theory deals with attaining reliable and efficient information transmission over a noisy channel.
The core of this subject lies in finding new error-correcting codes with improved parameters, along with developing their encoding and decoding algorithms.
The algebraic codes which possess interesting attributes are highly demanded in specific areas.
Codes which have some sort of randomness in their structure are admired by cryptographers.
The goal of coding theory is then to encode information in such a way that even if the channel (or storage medium) acquaint errors, the receiver can correct the errors and retrieve the original transmitted information. 
The term error-correcting code is used for both the detection and the correction mechanisms. 
In other words, to be accurate, we have error-detecting code and error-correcting code. 
The earlier one allows the detection of errors, whereas, later provides correction of errors discovered. 

Usually, coding is categorized as \textit{source coding} and \textit{channel coding}. Source coding involves changing the message source to a suitable code to be transmitted through the channel. 
An example of source coding is the ASCII code, which converts each character to a byte of 8 bits. 
The idea of channel coding is to encode the message again after the source coding by introducing some form of redundancy so that errors can be detected or even corrected.

This section covers the basic definitions and useful results which form grounds for Code-based cryptography. We begin by defining a code, a linear code, parity-check matrix, generator matrix, dimension of a linear code, Hamming distance of a code, Perfect codes, Syndrome decoding, Goppa codes, etc. While the underlying field upon which most codes rely, we keep discussing some points related to finite fields to give reader the complete understanding.


\subsection{Finite Fields}

Fields play a central role in algebra. For one thing, results about them find important applications in the theory of numbers.
The general theory of finite fields began with the work of Carl Friedrich Gauss ($1777–1855$) and Evaristé Galois ($1811–1832$), but it only became of interest for applied mathematicians and engineers in recent decades because of its many applications to mathematics, computer science and communication theory. 
Nowadays, the theory of finite fields has become very rich.

\begin{dfn}
	A nonempty set $\mathcal{R}$ is said to be a $ring$ if in $\mathcal{R}$ there are defined two operations, denoted by `$+$' (addition)  and `$\cdot$' (multiplication) respectively, such that for $a, b, c $ in $\mathcal{R}$:
	
	\begin{enumerate}[(i)]
		\item $a+b$ is in $\mathcal{R}$;
		\item $a+b = b+a$;
		\item $(a+b)+c = a+(b+c)$;
		\item There is an element $0$ in $\mathcal{R}$ such that $ a+0 =a$ (for every $a$ in $\mathcal{R}$);
		\item There exists an element `$-a$' in $\mathcal{R}$ such that $a + (-a)=0$;
		\item $a \cdot b$ is in $\mathcal{R}$;
		\item $a \cdot (b \cdot c) = (a \cdot b) \cdot c$;
		\item $a \cdot (b+c) = a \cdot b + a \cdot c$ and $(b+c)\cdot a = b \cdot a + c \cdot a$ (the two distributive laws).
	\end{enumerate}
	
	\textnormal{
		Axioms $(i)$ through $(iv)$ merely state that $\mathcal{R}$ is an abelian group under the operation `$+$', which we call addition. 
		Axioms $(vi)$ and $(vii)$ insist that $\mathcal{R}$ be closed under an associative operation `$\cdot$', which we call multiplication. 
		Axiom $(viii)$ serves to interrelate the two operations of $\mathcal{R}$.
		Further, if the multiplication of $\mathcal{R}$ is such that $a \cdot b = b \cdot a$ for every $a, b$ in $\mathcal{R}$, then $\mathcal{R}$ is said to be a} `commutative ring'.
\end{dfn}

From the notion of rings, we articulate the definition of fields which beholds the prime focus of this report. 
A definition of which by \textit{I.N.Hernstein} is stated as

\begin{dfn}
	
	Field is a commutative ring with unit element in which every nonzero element has a multiplicative inverse.
	
\end{dfn}

More generally, one may define a field by the following definition.

\begin{dfn}
	A \textit{field} is a nonempty set $F$ of elements with two operations `$+$' (addition) and `$\cdot$' (multiplication), which satisfies the following for all $a, b$ and $c \in F$:
	
	\begin{enumerate}[(i)]
		\item Closure w.r.t. addition and multiplication: $a+b \in F$ and $a \cdot b \in F$,
		\item Commutative w.r.t. addition and multiplication: $a+b=b+a$ and $a\cdot b = b \cdot a$,
		\item Associative w.r.t. addition and multiplication: $(a + b)+c = a + (b+c)$ and $(a \cdot b)\cdot c = a \cdot (b \cdot c)$,
		\item Distributivity holds on addition over multiplication: $ (a +b)\cdot c = a \cdot c + b \cdot c $.\vspace{1mm}\\
		Furthermore, for two elements $0$ and $1 \in F$ where $ 0 \neq 1$,
		\vspace{-1mm}
		\item $a+0 =a$ for all $a \in F$,
		\item $a \cdot 1 = a$ and $a \cdot 0 = 0$ for all $a \in F$,
		\item For any $a \in F$, there exist an additive inverse element `$-a$',
		\item For $a \neq 0$ in $F$, there exists a multiplicative inverse element $a ^{-1} \in  F$ such that $a \cdot a^{-1}=1$.  
	\end{enumerate}
\end{dfn}

\begin{example}
	\textnormal{
		Some common examples of fields include $\mathbb{R}$ (the set of real numbers), set of all rational numbers $\mathbb{Q}$, and the set of all complex numbers $\mathbb{C}$. The set of irrationals does not satisfy axiom $(i)$, hence it doesn't form a field, and the set of integers $\mathbb{Z}$ doesn't satisfy axiom $(viii)$, hence it is also not a field. Though, the set of integers `$modulo$' a prime number always form a field.}
\end{example}

\begin{dfn}
	Let F be a field. The characteristic of $F$ is the least positive integer `p' such that `$p\cdot1=0$', where $1$ is the multiplicative identity of $F$. If no such `$p$' exists, we define the characteristic to be $0$.
\end{dfn}

\begin{thm}
	The characteristic of a field is either $0$ or a prime number.
\end{thm}

Our main point of consideration would be fields which are finite. Finite fields have a specific algebraic structure and have cardinality as power of a prime. A finite field of order $2$ is denoted by $\F_2$ or $GF(2)$. 

\begin{thm}
	The set of integers `modulo m' i.e., $\mathbb{Z}_m$ forms a field if and only if `m' is a prime number.
\end{thm}

\begin{thm}
	Every finite field has order exactly a power of some prime number.
\end{thm}

\subsection{Polynomial Rings over Finite Fields}

Let $\mathbb{F}$ be a field; the set 
$$\mathbb{F}[x]:= \left\{ \sum_{i=0}^n a_i x^i : a_i \in \mathbb{F} \text{ for } 0 \leq i \leq n, \text{ and } n \geq 0  \right\}$$
along with polynomial addition and multiplication forms a ring and is called the \textit{polynomial ring} over $\mathbb{F}$. The elements of $\mathbb{F}[x]$ are called \textit{polynomials} over $\mathbb{F}$. The degree of a polynomial is the highest power of variable in it.

A polynomial $f(x)$ of positive degree is said to be reducible over $\mathbb{F}$ if there exist two polynomials $g(x)$ and $h(x)$ over $\mathbb{F}$ such that $1 \leq \deg g(x), \deg h(x) < \deg f(x)$ and $f(x) = g(x)h(x)$. If there does not exist such polynomials in $\F[x]$, then $f(x)$ is said to be irreducible over $\mathbb{F}$.

\begin{example}
	The polynomial $f(x) = x^4+ 2x^6 \in \mathbb{Z}_3[x]$ is of degree $6$ and, is reducible as $f(x)=x^4(1+2x^2)$. The polynomial $g(z)= 1 +z +z^2 \in \mathbb{Z}_2[z]$ is irreducible polynomial as there is no polynomial with acceptable degree as a factor of it. The polynomials having degree two or three are reducible if they have any root in the corresponding field.
\end{example}

We have the division algorithm, greatest common divisors, least common multiples etc. in these polynomial rings. Since for each $m >1$ of $\mathbb{Z}$, the ring $\mathbb{Z}_m = \mathbb{Z}/\left<m \right>$ is constructed, the similar relation holds for polynomial rings too. For $f(x) \in \mathbb{F}[x]$ having degree $n$, we define the fraction (quotient) ring

\begin{equation}
\frac{\mathbb{F}[x]}{\left<f(x)\right>} = \left\{ a_0 +a_1 x + a_2 x^2 + \cdots + a_{n-1} x^{n-1} + \left<f(x) \right> :\text{for } 0 \leq i \leq n-1, a_i \in \mathbb{F} \right\}.
\end{equation}

\begin{thm}
	Let $f(x)$ be a polynomial over a field $\mathbb{F}$ with degree $\geq 1$. Then the ring $\mathbb{F}[x]/\left<f(x)\right>$ is a field if and only if $f(x)$ is an irreducible polynomial over $\mathbb{F}$.
\end{thm}

\begin{example}
	The ring $\mathbb{Z}_2[x]/\left<1 + x + x^2 \right> = \{0,1,x, 1+x\}$ is a field of order $2^2 =4$.
\end{example}

Generally, for a $k$ degree irreducible polynomial $f(x) \in \mathbb{F}[x]$, where $\mid \mathbb{F} \mid = p$, the field $\mathbb{F}[x]/\left<f(x) \right>$ is of order $p^k$. A finite field of order $q$, denoted by $\F_q$, where $q= p^n$ for some prime $p$ and a natural number $n$ is described as:

$$\F_q= \frac{\mathbb{Z}_p[x]}{\left<f(x)\right>}$$
\begin{equation}
= \left\{a_0 + a_1 x + a_2 x^2 + \cdots + a_{n-1}x^{n-1} + \left< f(x) \right> : \text{ for }\textit{i} \in \{0,1,\dots,n-1\},  a_i \in \mathbb{Z}_p \right\}
\end{equation}

where $\mathbb{Z}_p[x]$ is a polynomial ring with variable $x$ and the coefficients from $\mathbb{Z}_p$, and $f(x)$ is an irreducible polynomial of degree $n$ over $\mathbb{Z}_p$.

An element $\alpha$ in a finite field $\mathbb{F}_q$ is called a \textit{primitive element} (or generator) of $\mathbb{F}_q$ if $\mathbb{F}_q=\{0, \alpha, \alpha^2, \dots , \alpha^{q-1}\}$. Consider the field $\mathbb{F}_4= \mathbb{F}_2[\alpha]$, where $\alpha$ is a root of the irreducible polynomial $1 + x + x^2 \in \mathbb{F}_2[x]$. Then we have 

\begin{center}
	\begin{tabular}{ll}
		$\alpha^2=$ & $-(1+\alpha)=1+\alpha$\\
		$\alpha^3=$ & $\alpha (\alpha^2) = \alpha(1+\alpha) = \alpha + \alpha^2 = \alpha + 1 + \alpha = 1.$
	\end{tabular}
\end{center}
From this, we see $\mathbb{F}_4 = \{0, \alpha, 1+\alpha, 1\}=\{0,\alpha,\alpha^2, \alpha^3\}$, so $\alpha$ is a primitive element of $\mathbb{F}_4$.

\begin{lem}
	The multiplicative order of any non-zero element $\alpha \in \mathbb{F}_q$ divides $q-1$. Further for any two non-zero elements $\alpha, \beta \in \mathbb{F}_q$, if $\gcd (|\alpha|, |\beta|)=1$, then $|\alpha \beta|= |\alpha|\times |\beta|$.
\end{lem}

\begin{thm}
	A non-zero element of $\mathbb{F}_q$ is a primitive element if and only if its multiplicative order is $q-1$; moreover, every finite field has at least one primitive element.
\end{thm}

Primitive elements are not unique for any field, in general. A \textit{minimal polynomial} of an element $\alpha \in \mathbb{F}_{q^m}$ with respect to $\mathbb{F}_q$ is a unique non-zero monic polynomial $f(x)$ of the least degree in $\mathbb{F}_q[x]$ such that $f(\alpha)=0$. The minimal polynomial is always irreducible over the base field; and the roots of this polynomial are all primitive elements of $\F_{q^m}$.

\begin{dfn}
	For $n$ being a natural number, $\F_q^n$ defines a set  $\{ (a_1,a_2,\dots,a_n): a_i \in \F_q \text{ for each } 1 \leq i \leq n \}$. Addition in $\F_q^n$ is coordinate-wise. This set forms an $n-$dimensional vector space over $\F_q$.
\end{dfn}

Let $\F_q$ be the finite field with $q$ elements. A non-empty set $V$, together with some (vector) addition $+$ and scalar multiplication by elements of $\F_q$,
is a \textit{vector space} over $\F_q$ if it satisfies the following conditions for all
$\textbf{u}, \textbf{v}, \textbf{w} \in V$ and for all $ \lambda, \mu \in \F_q$:
\begin{enumerate}[$(i)$]
	\item $\textbf{u} + \textbf{v} \in V$; \vspace{-2mm}
	\item $(\textbf{u} + \textbf{v}) + \textbf{w} = \textbf{u} + (\textbf{v} + \textbf{w})$;\vspace{-2mm}
	\item there is an element $\textbf{0} \in V$ with the property $\textbf{0} + \textbf{v} = \textbf{v} + \textbf{0}$;\vspace{-2mm}
	\item for each $\textbf{u} \in V$ there is an element of $V$, called $-\textbf{u}$, such that $\textbf{u} + (-\textbf{u}) = 0 = (-\textbf{u}) + \textbf{u}$;\vspace{-2mm}
	\item $\textbf{u}+\textbf{v}=\textbf{v}+\textbf{u} $;
	\item $\lambda \textbf{v} \in V$;\vspace{-2mm}
	\item $\lambda (\textbf{u}+\textbf{v})= \lambda \textbf{u} + \lambda\textbf{v}$;\vspace{-2mm}
	\item $(\lambda + \mu)\textbf{u}= \lambda \textbf{u}+\mu \textbf{u}$;\vspace{-2mm}
	\item $(\lambda \mu)\textbf{u}= \lambda (\mu \textbf{u})$;\vspace{-2mm}
	\item if $1$ is the multiplicative identity of $\F_q$, then $1 \textbf{u} = \textbf{u}$.
\end{enumerate}

In the following section, the term `messages' will be a tuple of certain fixed size length with entries from specified set/ field.

\subsection{Basic Coding Theory}

The objective of Coding Theory is the transmission of messages over noisy channels. The basic visualization is designed as below:

\begin{center}
	\includegraphics[width= 13.5cm, height=4cm]{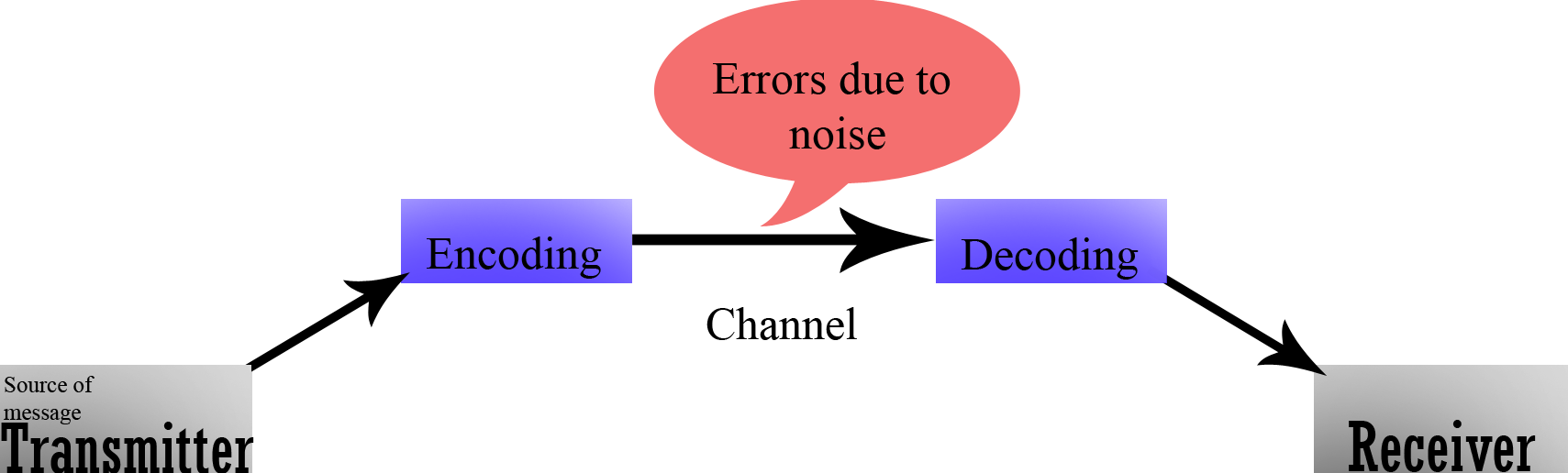} \\
	\textbf{Figure 1:} Transmission over noisy channel
\end{center}

From this we see that if message $\textbf{m}$ is transmitted and $\textbf{e}$ is the error that occurred during transmission of $\textbf{m}$ and $\textbf{y}$ is received by the receiver, then we have $\textbf{y} = \textbf{x} + \textbf{e}$. Here $`+$' is coordinate-wise addition.

\begin{dfn}
	
	A \textbf{block code}  $\mathcal{C}$ of length $n$ is a subset of $\mc{A}^n$, where $\mc{A}$ is said to be the set of alphabets of $\mathcal{C}$. Generally, $\mc{A}$ is a finite field. A \textbf{q-ary block code} of length n is a given set of sequences of length n, of symbols where each symbol is chosen from a finite field $\F_q$. A block code of length $n$ in which every codeword is repitition of a single symbol is called a \textbf{repetition code} of length $n$.
	
\end{dfn}

\begin{dfn}
	The \textbf{Hamming distance} between any two vectors of $\F_q^n$, denoted by $d_H(\textbf{x},\textbf{y}):=$ Number of differences in digits of $\textbf{x}$ and $\textbf{y}$.
	The \textbf{Hamming weight} $wt(\textbf{x})$ of the bit string $\textbf{x} \in \F_2^n$ is the number of nonzero coordinates in $\textbf{x}$. The relation between distance and weight is $wt(\textbf{x})=d(\textbf{x},\textbf{0})$. The \textbf{Hamming distance of a code} $\mathcal{C}$ is defined as $d(\mathcal{C}):= min\{d(\textbf{x},\textbf{y}) \text{ for all } \textbf{x}, \textbf{y} \in \mathcal{C}, \text{ with } \textbf{x} \neq \textbf{y} \}$.	
	
\end{dfn}

For every element $x$ of $\F_q$, we can define Hamming weight as follows:
$$wt(x) = d(x,0)= \left\{ \begin{matrix}
1 & \text{ if }x \neq 0\\
0 & \text{ if }x =0.
\end{matrix}\right.$$

\begin{thm}
	
	The Hamming distance defines a metric on $\F_q^n$. If $\textbf{x}, \textbf{y} \in \F_q^n$, then $d(\textbf{x},\textbf{y})=wt(\textbf{x}-\textbf{y})$. 
	As a consequence of this, $\textbf{x}$ and $\textbf{y}$ being vectors in $\F_2^n$, $d_H(\textbf{x},\textbf{y})=wt(\textbf{x}+\textbf{y})$. Equivalently, let $q$ be even. If $\textbf{x},\textbf{y} \in \F_q^n$, then $d(\textbf{x}, \textbf{y})= wt (\textbf{x}+\textbf{y})$. Furthermore, if $\textbf{x},\textbf{y} \in \F_2^n$, then $wt (\textbf{x}+\textbf{y})= wt (\textbf{x})+wt(\textbf{y})- 2 wt (\textbf{x} \star \textbf{y})$, where $\star$ represents component-wise multiplication.
	
\end{thm}

\begin{dfn}
	Encoding is defined as a function
	`$Encode: \{ \text{Messages set} \} \longrightarrow \text{Code} \subseteq \mc{A}^n$'.
\end{dfn}

The Encoding function is not common in general. For an instance, suppose we have two messages: $Yes$ and $No$, then we can encode $Yes \rightarrow 1$ and $No \rightarrow 0$; but, in this case due to noise in channel if we intended to transmit the message $1$ and received $0$ in response, there would be no clue to receiver that the original message was $1$. Considering the encoding based on majority bits decoding: $Yes \rightarrow 111$ and $No \rightarrow 000$ then upon error in transmission up to one place in channel, the receiver can easily detect and correct the error. In latter case, the code used is \textit{binary repetition code of length 3 over $\F_2$}. In this case the decoded message is the majority bit. This defines the general problem of Coding Theory: \textit{Error Detection and Correction}.

%
%


\begin{thm} 
	A code $\mathcal{C}$ can detect upto `s' errors in any codeword if $d(\mathcal{C}) \geq s+1$ and it can correct upto `t' errors in any codeword if $d(\mathcal{C})\geq 2t+1$.
\end{thm}

%
%
%
%
%
%
%
%
%
%
%
%
%
%
%
%
%
%
%
%
%
%
%
%
%
%
%
%
%
%

\begin{dfn}[Sphere]
	
	A sphere of radius {$r$}, centred about $\textbf{u} \in \F_q^n$ is defined as: $$S(\textbf{u},r):= \{ \textbf{v} \in \F_q^n: d(\textbf{u},\textbf{v}) \leq r\}.$$
	
\end{dfn}

\begin{thm}
	
	For some $\textbf{u} \in \F_q^n$ and $r\in \{0, 1, \dots, n\}$, $S(\textbf{u},r)$ contains exactly $ {n\choose0} + {n\choose1}  (q-1) + {n\choose2}  (q-1)^2 + \cdots + {n\choose r}  (q-1)^r$ elements of $\F_q^n$.
	
\end{thm}

\begin{thm}[Sphere Packing or Hamming Bound]
	
	If $M$ is the number of codewords in a code $\mc{C}$ having length `$n$' over $\F_q$ with $d(\mc{C}) \geq 2t+1$, then we have: 
	
	$$ M\left( {n\choose0} + {n\choose1}  (q-1) + {n\choose2}  (q-1)^2 + \cdots + {n\choose t}  (q-1)^t\right) \leq q^n.$$
	
\end{thm}

\begin{dfn}[Perfect Code]
	
	A code which satisfies equality in the sphere packing bound.
	
\end{dfn}

\begin{example}
	
	Following are some examples of perfect codes.
	
	\begin{enumerate}[(i)]
		
		\item Binary repetition code of length 5,
		
		\item Code consisting all elements of $\F_q^n$.
		
	\end{enumerate}
	
\end{example}

\begin{thm}[Plotkin construction : ($\textbf{u}|\textbf{u}+\textbf{v}$)]
	
	If $\mc{C}_1$ be some binary code of length $n$, having $M_1$ codewords, with $d(\mathcal{C}_1)= d_1$, and $\mc{C}_2$ be some binary code of length $n$, having $M_2$ codewords, with $d(\mathcal{C}_2)= d_2$, then the binary code $\mc{C}_3:= \{(\textbf{u}\mid \mid\textbf{u}+\textbf{v}): \textbf{u} \in \mc{C}_1, \textbf{v} \in \mc{C}_2 \}$ is a binary code of length $2n$, with $M_1 M_2$ codewords, and $d(\mc{C}_3) = \min \{2d_1, d_2 \}$. 
	
\end{thm}

If we suppose for any natural number $n$, $E_n$ be the set of vectors in $\F_2^n$ which have even weight. This, together with Plotkin construction produces an interesting family of codes:

We begin by choosing $\mc{C}_1$ to be $E_4$ and $\mc{C}_2$ to be binary repetition code of length $4$, then using Plotkin construction we get $\mc{C}_3$ and extending this type of construction with a binary repetition code of length $8$. By doing this construction time and again we get codes with length $2^m$, having $2^{m+1}$ codewords, and distance $2^{m-1}$ for $m\geq2$. These codes are known as the first-order $Reed-Muller$ codes.

\begin{thm}
	
	$Reed-Muller$ first order codes are optimal with those parameters, $i.e.,$ these codes achieve Sphere-packing bound.
	
\end{thm}

The $ISBN$ (International Standard Book Number), a $10$ digit number, say $x_1 x_2 x_3 \cdots x_{10}$ is designed in such a way that it satisfies $\sum_{i=1}^{10} i x_i \equiv 0 (mod \, 11)$. This is called \textit{Weighted check sum}. Using this, one can detect if there is an error or not in $ISBN$.


\subsection{Linear Codes}

Linear codes of length $n$ over $\F_q$ are the subspaces of $\F_q^n$, thus the dimension of these subspaces is finite. Since a linear code is a vector space, all its elements can be described in terms of a basis. We first recall some facts from linear algebra.
Knowing a basis for a linear code enables us to describe its codewords explicitly. 
In coding theory, a basis for a linear code is often represented in the form of a matrix, called a generator matrix, while a matrix that represents a basis for the dual code is called a parity-check matrix. 
These matrices play an important role in coding theory.

\begin{dfn}[$\textbf{q} - $ary linear code]
	Let $\mc{C}$ be a $k - $dimensional subspace of $\F_q^n$ then we say that $\mc{C}$ is a $q$-ary $[n, k]$ code; and if $d(\mc{C})=d$, then $\mc{C}$ is a $q$-ary $[n,k,d]$ code over $\F_q$.
\end{dfn}

\begin{lem}
	A $q - $ary $[n,k,d]$ code contains $q^k$ codewords. If $\textbf{x} \text{ and } \textbf{y} \in \F_q^n$, then $d(\textbf{x},\textbf{y})=w(\textbf{x} - \textbf{y})$. 
\end{lem}

\begin{thm}
	Let $\mc{C}$ be a linear code and $w(\mc{C})$ be the smallest of weights out of the non-zero codewords of $\mc{C}$, then $d(\mc{C})=w(\mc{C})$.
\end{thm}

\begin{dfn}[Generator Matrix]
	The {Generator Matrix} for a $q - $ary linear $[n,k]$ code $\mc{C}$ is a $k\times n$ matrix, whose rows form the basis of vector space $\mc{C}$ over $\F_q$.
\end{dfn}

\begin{dfn}[Parity Check Matrix]
	Let $\mc{C}$ be a $q - $ary linear $[n,k]$ code with generator matrix $G$, then a $(n-k)\times n$ matrix $H$ is called a Parity Check matrix for $\mc{C}$ if $G \cdot H^T=0$.
\end{dfn}

If  $\mc{C}$ is an $[n,k]-$linear code, then a generator matrix for $\mc{C}$ must be a $k\times n$ matrix and a parity-check matrix for $\mc{C}$ must be an $(n-k)\times n$ matrix.

As the number of bases for a vector space usually exceeds one, the number of generator matrices for a linear code also usually exceeds one. 
Moreover, even when the basis is fixed, a permutation (different from the identity) of the rows of a generator matrix also leads to a different generator matrix.

The rows of a generator matrix are linearly independent. 
The same holds for the rows of a parity-check matrix. 
To show that a $k \times n$ matrix $G$ is indeed a generator matrix for a given $[n,k]-$linear code $\mc{C}$, it suffices to show that the rows of $G$ are codewords in $\mc{C}$ and that they are linearly independent. 
Alternatively, one may also show that $\mc{C}$ is contained in the row space of $G$.

%
%
%
%
%
%
%
%

The family of linear codes having same parameters are divided into various Equivalence classes.

\begin{dfn} 
	Any pair of linear codes are Equivalent if the generator matrix of one can be obtained from other by either or the combination of the following:
	\begin{itemize}
		\item Permutation of rows,
		\item Multiplication of rows by a non-zero scalar,
		\item Addition of a scalar multiple of one row to another,
		\item Permutation of columns,
		\item Multiplication of any column by a non-zero scalar.
	\end{itemize}
\end{dfn}

As we have divided the family of linear codes into equivalence classes, we now require a class representative.	Supposing $G$ to be the generator matrix for $[n,k]$-code $\mc{C}$, on applying equivalence criteria, $G$ can be transformed to standard form: $$G_{k\times n}= \left[I_k \mid A_{k \times (n-k)}\right]_{k\times n}.$$

\begin{dfn}
	A generator matrix of the form $(I_k \mid X)$ is said to be in standard form. A parity-check matrix in the form $(Y \mid I_{n-k})$ is said to be in standard form.
\end{dfn}

If $G = (I_k \mid X)$ is the standard form generator matrix of an $[n,k]-$code $\mc{C}$, then a parity-check matrix for $\mc{C}$ is $H = (-X^T \mid I_{n-k})$.
It should be noted that it is not true that every linear code has a generator matrix in standard form.
The standard form for any code is not unique; it is unique up to Equivalence of codes.

\begin{example}
	\textnormal{The code $\mc{C} = \{000,001,100,101\}$ is a $2-$dimensional code and its generator matrix do not have a standard form.}
\end{example}

\begin{lem}
	Let $\mc{C}$ be an $[n,k]-$linear code over $\F_q$, with generator matrix $G$. 
	Then $\textbf{v} \in \F_q^n$ belongs to $\mc{C}^{\perp}$ if and only if $\textbf{v}$ is orthogonal to every row of $G$; $i.e.,$ $v \in \mc{C}^{\perp} \iff \textbf{v}G^T = \textbf{0}$. 
	In particular, given an $(n-k)\times n$ matrix $H$, then $H$ is a parity-check matrix for $\mc{C}$ if and only if the rows of $H$ are linearly independent and $HG^T = 0$.
\end{lem}

\begin{thm}
	Let $\mc{C}$ be a linear code and let $H$ be a parity-check matrix for $\mc{C}$. Then
	\begin{enumerate}[(i)]
		\item $\mc{C}$ has distance $\geq d$ if and only if any $d-1$ columns of $H$ are linearly independent; and
		\item $\mc{C}$ has distance $\leq d$ if and only if $H$ has $d$ columns that are linearly dependent.
	\end{enumerate}
\end{thm}

\begin{cor}
	Let $\mc{C}$ be a linear code and let $H$ be a parity-check matrix for $\mc{C}$. 
	Then the following statements are equivalent:
	\begin{enumerate}[(i)]
		\item $\mc{C}$ has distance $d$;
		\item any $d-1$ columns of $H$ are linearly independent and $H$ has $d$ columns that are linearly dependent.
	\end{enumerate}
\end{cor}

\begin{example}
	
	$E_n$, being a linear code, is a $[n,n-1,2]$-code over $\F_2$, with  generator matrix in standard form as: 
	$$\left[ 
	\begin{array}{ccccc|c}
	1 & 0& 0 & \cdots & 0 & 1 \\ 
	0 & 1& 0 & \cdots & 0 & 1 \\ 
	0 & 0& 1 & \cdots & 0 & 1 \\ 
	\vdots& \vdots & \vdots &\ddots & \vdots & \vdots \\
	0 & 0& 0 & \cdots & 1 & 1\\ 
	\end{array}
	\right] $$
	
\end{example}

\begin{example} 
	
	For $H$ being an $r \times n$ matrix over $\F_q$, the null space of linear transformation corresponding to $H$ is a linear $[n,nullity(H)]$-code over $\F_q$. The matrix $H$ is then the {parity-check matrix} for this code.
	
\end{example}

%
%
%
%
%
%
%
%
%
%
%
%
%
%
%
%
%
%

\subsection{Encoding and Decoding with Linear Codes}

Coding theory offers versatile linear codes which are employed to cater encoding and decoding of messages. These can be regarded as invertible functions over vector spaces. The encoding and decoding processes are described in the remaining component of this section. 

\subsection{Encoding Process}

Let $\mc{C}$ be an $[n,k,d]-$linear code over the finite field $\F_q$. Each codeword of $\mc{C}$ can represent one piece of information, so $\mc{C}$ can represent $q^k$ distinct pieces of information.
Suppose $\textbf{m}=(m_1, m_2,\dots, m_k)$ is a message in $\F_q^k$, then by encoding, we mean a one-to-one function $$\text{Encode} : \F_q^k \longrightarrow \F_q^n$$
$i.e.,$ we append `$n-k$' extra information bits to message $\textbf{m}$ as  $$(m_1, m_2,\dots, m_k) \mapsto (\hat{m_1}, \hat{m_2},\dots, \hat{m_k}, \underbrace{\hat{m_{k+1}}, \dots, \hat{m_n}}_{n-k \text{ redundancy bits}}).$$

Let $\mc{C}$ be an $[n,k]$-code over $\F_q$ with generator matrix $G$. Then $\mc{C}$ contains $q^k$ codewords and thus $\mc{C}$ can be used to communicate $q^k$ number of messages at a time.
Encoding with a linear code can be done by multiplication of message vector with generator matrix of the linear code as $\textbf{m} \mapsto \textbf{m} \cdot G$.
The end vector $\textbf{m} \cdot G$ is indeed a codeword of $\mc{C}$, as it is a linear combination of the rows of generator matrix.
Encoding map is simpler when the generator matrix is given in standard form as $[I_k|A_{k \times (n-k)}]$.\\

Suppose the codeword $\textbf{x}=(x_1,x_2,\dots,x_n)$ is sent through the channel and that the vector $\textbf{y}=(y_1,y_2,\dots,y_n)$ is received. We define \textit{error vector} \textbf{e} as: $$\textbf{e}=\textbf{y}-\textbf{x} = (e_1,e_2,\dots,e_n).$$
The decoder must decide from \textbf{y} which codeword \textbf{x} was transmitted, or equivalently what is corresponding error vector \textbf{e}. This process is called decoding with a linear code, and is achieved with various methods: nearest neighbor decoding, syndrome decoding, standard array decoding, etc.

We begin decoding by first looking at notion of a coset. Cosets play an essential role in many decoding schemes.

\begin{dfn}
	Let $\mc{C}$ be a linear code of length $n$ over $\F_q$, and let $\textbf{u} \in \F_q^n$ be any vector of length $n$; we define the coset of $\mc{C}$ determined by $\textbf{u}$ to be the set 
	$$\mc{C}+ \textbf{u} = \{ \textbf{v} + \textbf{u}: \textbf{v}\in \mc{C} \} = \textbf{u} + \mc{C}.$$
\end{dfn}

\begin{thm}
	Let $\mc{C}$ be an $[n,k,d]-$linear code over the finite field $\F_q$. Then,
	\begin{enumerate}[(i)]
		\item every vector of $F_q^n$ is contained in some coset of $\mc{C}$; 
		\item for all $\textbf{u} \in \F_q^n$, $|\mc{C} +\textbf{u}|=|\mc{C}|=q^k;$ 
		\item for all $\textbf{u},\textbf{v}\in \F_q^n, \textbf{u} \in \mc{C} + \textbf{v}$ implies that $\mc{C} +u=\mc{C} +\textbf{v}$; 
		\item two cosets are either identical or they have empty intersection; 
		\item there are $q^{n-k}$ different cosets of $\mc{C}$; 
		\item for all $\textbf{u},\textbf{v}\in \F_q^n, \textbf{u}-\textbf{v} \in \mc{C}$ if and only if $\textbf{u}$ and $\textbf{v}$ are in the same coset. 
	\end{enumerate}
\end{thm}

\begin{dfn}
	A vector in a coset is called a coset leader if it has the minimum Hamming weight.
\end{dfn}

\subsection{Nearest Neighbor Decoding}

Let $\mc{C}$ be a linear code. Assume the codeword $\textbf{v}$ is transmitted and the vector $\textbf{w}$ is received, resulting in the error vector $\textbf{e}=\textbf{w}-\textbf{v} \in \textbf{w} + \mc{C}$. 
Then $\textbf{w} - \textbf{e} = \textbf{v} \in \mc{C}$ by axiom $(vi)$, so the error vector $\textbf{e}$ and the received vector $\textbf{w}$ are in the same coset.
Since error vectors of small Hamming weight are the most likely to occur, nearest neighbor decoding works for a linear code $\mc{C}$ in the following manner. 
Upon receiving the vector $\textbf{w}$, we choose a vector $\textbf{e}$ of least Hamming weight in the coset $\textbf{w}+\mc{C}$ and conclude that $\textbf{v}=\textbf{w}-\textbf{e}$ was the codeword transmitted.
This process is executed as:

Let $\mc{C}=\{c_1,c_2,\cdots,c_k\}$ be a linear code over $\F_q$ of length $n$, and $G$ be its generator matrix. A \textit{standard array} matrix is formed for the code $\mc{C}$ as follows:

\begin{enumerate}[Step $(i)$:]
	
	\item The first row of the matrix consists of the elements of $\mc{C}$,
	
	\item Find a smallest weighted vector of $\F_q^n$ which do not lie in first row (select any if more than one) say $a_1$, then the second row is the elements of the coset $a_1 +\mc{C}$,
	
	\item Repeat above process, taking the smallest weighted vector not lying in above rows till all the elements of $\F_q^n$ are exhausted.
	
\end{enumerate}

This matrix will consist all the elements of $\F_q^n$. 
Taking element of $\F_q^k$ as input, its encoding is done by post multiplication with $G$.
For decoding process, suppose the received vector is $a_l+c_w$ (looking in standard array), then its decoding will be $c_w$ and $a_l$ is the error.

\begin{example}
	
	Let $\mc{C}=\left< 1011, 0101 \right>$ be a linear code over $\F_2^4$, then the standard array of $\mc{C}$ is\\ 
	
	$\mc{C}= 
	\left\{
	\begin{array}{cccc}
	
	0000 & 1011 & 0101 & 1110 \\
	1000 & 0011 & 1101 & 0110\\
	0100 & 1111 & 0001 & 1010\\
	0010 & 1001 & 0111 & 1100\\ 
	
	\end{array} 
	\right. $ \\ 
	
	So if $0001$ is received, the decoded message to this is $0101$ with error vector $0100$.
\end{example}

If the code has greater parameters, more computations would be required, which increases the complexity, hence we have Syndrome decoding.

\subsection{Syndrome Decoding}

The decoding scheme based on the standard array works reasonably well when the length $n$ of the linear code is small, but it may take a considerable amount of time when $n$ is large. 
Time can be saved by making use of the syndrome to identify the coset to which the received vector belongs.
For this, we need to understand a few things first.

\begin{dfn}[Dual of a code]
	Given a linear code $\mc{C}$ over $\F_q$, the dual code of $\mc{C}$ is defined as $$\mc{C}^{\perp}:= \{\textbf{v} \in \F_q^n : \textbf{u} \cdot \textbf{v} = 0 \text{ for all } \textbf{u} \in \mc{C} \}.$$
\end{dfn}

The dot product used here is standard Euclidean dot product.

\begin{thm} 
	The dual code of any linear code is also linear code over the same field. Further, if $\mc{C}$ is $[n,k]$-code over $\F_q$, then $\mc{C}^{\perp}$ is $[n,n-k]$-code over $\F_q$; also, $\left(\mc{C}^{\perp}\right)^{\perp}=\mc{C}$
\end{thm}

\begin{dfn} 
	A parity-check matrix $H$ for an $[n,k]$-code $\mc{C}$ is a generator matrix for $\mc{C}^{\perp}$.
\end{dfn}

\begin{thm}
	If $G= \left[ I_k | A_{k \times (n-k)} \right]_{k \times n}$ is a generator matrix for some $[n,k]$-code $\mc{C}$, then the generator matrix for $\mc{C}^{\perp}$ is $H= \left[ -A^T_{(n-k)\times k} | I_{n-k} \right]_{(n-k)\times n}.$ This is the standard form of parity-check matrix for $\mc{C}$.
\end{thm}

\begin{dfn}[Syndrome of a vector] 
	Let $\mc{C}$ be an $[n,k,d]-$linear code over $\F_q$ and let $H$ be a parity-check matrix for $\mc{C}$. For any $\textbf{w} \in \F_q^n$, the syndrome of $\textbf{w}$ is the word $S(\textbf{w}) = \textbf{w}H^T \in \F_q^{n-k}$. (Strictly speaking, as the syndrome depends on the choice of the parity-check matrix $H$, it is more appropriate to denote the syndrome of $\textbf{w}$ by $S_H(\textbf{w})$ to emphasize this dependence. However, for simplicity of notation, the subscript $H$ is dropped whenever there is no risk of ambiguity.)
\end{dfn}

\begin{rem} Let $\mc{C}$ be an $[n,k,d]-$linear code and let $H$ be a parity-check matrix for $\mc{C}$. For $\textbf{u},\textbf{v}\in \F_q^n$, we have 		
	\begin{enumerate}[(i)]
		\item $S(\textbf{u}+\textbf{v})= S(\textbf{u})+S(\textbf{v})$;
		\item $S(\textbf{y})=\textbf{0}$ if and only if $\textbf{y} \in \mc{C}$;
		\item $S(\textbf{u})= S(\textbf{v})$ if and only if $\textbf{u}$ and $\textbf{v}$ are in the same coset of $\mc{C}$.
	\end{enumerate} 
\end{rem}

A table which matches each coset leader with its syndrome is called a syndrome look-up table. 

Steps to construct a \textit{Syndrome look-up table} assuming complete nearest neighbor decoding:

\begin{enumerate}[Step $(i)$:]
	\item List all the cosets for the code, choose from each coset a word of least weight as coset leader $\textbf{u}$.
	\item Find a parity-check matrix $H$ for the code and, for each coset leader $\textbf{u}$, calculate its syndrome $S(\textbf{u})=\textbf{u}H^T$.
\end{enumerate}

The Syndrome decoding works in same way as standard array decoding works. The Syndrome of received vector is calculated and the coset leader of the row in which it lies is found by matching the syndrome of coset leaders. This reduces the time of searching the row of standard array. The decoding procedure for Syndrome decoding is as:

\begin{enumerate}[Step $(i)$:]
	\item For the received vector $\textbf{w}$, compute the syndrome $S(\textbf{w})$.
	\item Find the coset leader $\textbf{u}$ next to the syndrome $S(\textbf{w}) = S(\textbf{u})$ in the syndrome look-up table.
	\item Decode $\textbf{w}$ as $\textbf{v}= \textbf{w} - \textbf{u}$.
\end{enumerate}

Unlike the above stated codes, we now focus on codes which are formulated and designed from another crucial expect of encoding, which is the number of errors they can correct. Therefore, in next section, we describe the algebraic structure of Goppa codes in which the error correcting capability corresponds to the degree of the Goppa polynomial we select.

\subsection{Goppa Codes}

Born in $1939$, a Soviet and Russian mathematician, \textit{Valery Denisovich Goppa}, discovered the relation between algebraic geometry and codes in $1970$. This led to the idea of \textit{Goppa Codes}.  
It turned out that Goppa codes also form arguably the most interesting subclass of alternant codes, introduced by H. J. Helgert in 1974. 
These codes have got efficient decoding algorithm by N. Patterson \cite{patterson} in $1975$.

\begin{dfn}
	Let $g(z)=g_0 + g_1 z + g_2 z^2 +\cdots + g_t z^t \in \F_{q^m} [z]$, and let $L=\{\alpha_1,\alpha_2 ,\dots,\alpha_n\} \subseteq \F_{q^m}$ such that, $g(\alpha_i) \neq 0, \text{ for all } \alpha_i \in L$. Then the code defined by $$ \left\{ \textbf{c}=(c_1,c_2,\dots,c_n) \in \F_q^n :\sum_{i=1}^n \frac{c_i}{z-\alpha_i} \equiv 0 \,\mod g(z) \right\}$$ is called $\textbf{Goppa code}$ with parameters $g(z)$ and $L$; denoted by $\Gamma(L,g(z))$. 
\end{dfn}

For each $i$  $(where \, 1 \leq i \leq n)$, $g(\alpha_i) \neq 0$ equivalently $\gcd(z-\alpha_i,g(z))=1$, the fraction $\frac{1}{z-\alpha_i}$ is computed in $\frac{\F_{q^m} [z]}{\left<g(z)\right>}$ as

\begin{thm}\label{one}
	The multiplicative inverse of $(z-\alpha_i)$ exists in the quotient ring $\frac{\F_{q^m} [z]}{\left<g(z)\right>}$; the value of $(z-\alpha_i)^{-1}$ in $\frac{\F_{q^m} [z]}{\left<g(z)\right>}$ is $-\left( \frac{g(z) -g(\alpha_i)}{z-\alpha_i} \right) g(\alpha_i)^{-1}$. A vector $\textbf{c} \in \Gamma(L,g)$ if and only if $\sum_i c_i \left( \frac{g(\alpha_i)-g(z)}{z-\alpha_i} \right)g(\alpha_i)^{-1} \equiv 0 \text{ (mod \textit{g(z)})}$. 
\end{thm}

Using this result, we can derive the following most important corollary:

\begin{cor}
	A vector $\textbf{c} \in \Gamma(L,g)$ if and only if $\sum_i c_i \left( \frac{g(\alpha_i)-g(z)}{z-\alpha_i} \right)g(\alpha_i)^{-1} =0$ as a polynomial in $\F_{q^m}[z]$. 
\end{cor}

Hence, we derive the Parity check matrix over $\F_{q^m}$ for the Goppa codes as:

\begin{cor}
	For a Goppa code $\Gamma(L,g(z))$, the Parity check matrix over $\F_{q^m}$ is $H$  $$= 
	\left[ \begin{array}{cccc}
	g_t g(\alpha_1)^{-1} & g_t g(\alpha_2)^{-1} & \cdots & g_t g(\alpha_n)^{-1}  \\
	(g_t \alpha_1 + g_{t-1})g(\alpha_1)^{-1} & (g_t \alpha_2 + g_{t-1})g(\alpha_2)^{-1} & \cdots & (g_t \alpha_n + g_{t-1})g(\alpha_n)^{-1} \\
	\vdots & \vdots & \vdots & \vdots \\
	(g_t \alpha_1^{t-1}  + \cdots + g_1)g(\alpha_1)^{-1} & (g_t \alpha_2^{t-1}  + \cdots + g_1)g(\alpha_2)^{-1} & \cdots & (g_t \alpha_n^{t-1}  + \cdots + g_1)g(\alpha_n)^{-1} \\
	\end{array} \right].$$
\end{cor}

This matrix is further broken down into product of three matrices as:
$$H = \underbrace{\left[ \begin{array}{cccc}
	g_t & 0 & \cdots & 0 \\
	g_{t-1} & g_t & \cdots & 0  \\
	\vdots & \vdots & \vdots & \vdots \\
	g_1 & g_2 & \cdots & g_t \\ \end{array} \right]}_{C}  \underbrace{\left[ \begin{array}{cccc}
	1 & 1 & \cdots & 1 \\
	\alpha_1 & \alpha_2 & \cdots & \alpha_n  \\
	\vdots & \vdots & \vdots & \vdots \\
	\alpha_1^{t-1} & \alpha_2^{t-1} & \cdots & \alpha_n^{t-1} \\ \end{array} \right]}_X  \underbrace{\left[ \begin{array}{cccc}
	g(\alpha_1)^{-1} & 0 & \cdots & 0 \\
	0 & g(\alpha_2)^{-1} & \cdots & 0  \\
	\vdots & \vdots & \vdots & \vdots \\
	0 & 0 & \cdots & g(\alpha_n)^{-1} \\ \end{array} \right]}_Y.$$

Now, as we have $\textbf{c} \in \Gamma(L,g)$ if and only if $\textbf{c}H^T=0$, which implies $\textbf{c}(CXY)^T=0$, equivalently $\textbf{c}Y^T X^T C^T=0$, and this gives $\textbf{c}Y^T X^T=0$, or $\textbf{c}(XY)^T=0$ (as matrix $C$ is invertible).

\begin{rem}\label{parmat}
	The matrix $XY$ can be viewed as Parity check matrix for $\Gamma(L,g)$ over $\F_{q^m}$. The matrix $$XY =\left[ \begin{array}{cccc}
	g(\alpha_1)^{-1} & g(\alpha_2)^{-1} & \cdots & g(\alpha_n)^{-1} \\
	\alpha_1 g(\alpha_1)^{-1} & \alpha_2 g(\alpha_2)^{-1} & \cdots & \alpha_n g(\alpha_n)^{-1}  \\
	\vdots & \vdots & \vdots & \vdots \\
	\alpha_1^{t-1} g(\alpha_1)^{-1} & \alpha_2^{t-1} g(\alpha_2)^{-1} & \cdots & \alpha_n^{t-1} g(\alpha_n)^{-1} \\ \end{array} \right]_{t \times n}.$$
\end{rem}

\begin{rem}
	Viewing elements of $\F_{q^m}$ as vectors of length `$m$' over $\F_q$ by vector space isomorphism, we have a Parity check matrix for $\Gamma(L,g)$ over $\F_q$ to be an `$mt \times n$' matrix, with at least {`$t$'} columns linearly independent over $\F_q$. Hence, the Hamming distance of Goppa Code, $d(\Gamma(L,g)) \geq t +1$.
	Since, for the matrix $XY$ over $\F_q$, maximum of {`$mt$'} rows are linearly independent, hence, Rank$(XY) \leq mt$, which gives Nullity$(XY) \geq n-mt$. Therefore, dimension of Goppa code, {$dim_{\F_q} \Gamma(L,g) \geq n-mt$}.
\end{rem}

\begin{dfn}[Primitive Polynomial]
	An irreducible polynomial `{$p(x)$}' of degree `{$m$}' over $\F_q$ is called primitive polynomial if its roots form primitive elements of $\F_{q^m}$. 
\end{dfn}
For example $p(x)= x^{13}+x^4+x^3+x+1$ is a primitive polynomial of degree 13 over binary field $\F_2$.


\begin{rem}
	There are $\frac{\phi(q^n-1)}{n}$ primitive polynomials of degree `$n$' over $\F_q$. In construction of Goppa codes, the extension field is constructed taking modulo a primitive polynomial.
\end{rem}

\begin{example}[A Goppa Code]{\label{example}}
	Let $\F_{2^4}$ be the field isomorphic to $\frac{\F_2[x]}{\left<x^4+x+1\right>}$. Let `$\alpha$' be a root of $x^4+x+1$, then, since multiplicative order of `$\alpha$' is $15$, it can be used to generate all elements of $\F_{2^4}^*$; equivalently, $\F_{2^4}^* = \left< \alpha \right>$.
	Hence we can represent elements of $\F_{2^4}$ as 
	\begin{center}
		$ \begin{array}{ccllllc}
		0&=&&&&&=(0,0,0,0)^T;\\
		1& =&  1 & & & & = (1,0,0,0)^T;\\
		\alpha & = & &\alpha & & & = (0,1,0,0)^T;\\
		\alpha^2 & = & & &\alpha^2 & & = (0,0,1,0)^T;\\
		\alpha^3 & = & & & &\alpha^3 & = (0,0,0,1)^T;\\
		\alpha^4 & = &1+ &\alpha & & & = (1,1,0,0)^T;\\
		\alpha^5 & = & &\alpha+ &\alpha^2 & & = (0,1,1,0)^T;\\
		\alpha^6 & = & & &\alpha^2+ &\alpha^3 & = (0,0,1,1)^T;\\
		\alpha^7 & = & 1+&\alpha+ & & \alpha^3& = (1,1,0,1)^T;\\
		\alpha^8 & = & 1+& &\alpha^2 & & = (1,0,1,0)^T;\\
		\alpha^9 & = & &\alpha+ & &\alpha^3 & = (0,1,0,1)^T;\\
		\alpha^{10} & = &1+ &\alpha+ & \alpha^2& & = (1,1,1,0)^T;\\
		\alpha^{11} & = & & \alpha+&\alpha^2+ & \alpha^3& = (0,1,1,1)^T;\\
		\alpha^{12} & = &1+ & \alpha+&\alpha^2+ & \alpha^3& = (1,1,1,1)^T;\\
		\alpha^{13} & = & 1+& & \alpha^2+& \alpha^3& = (1,0,1,1)^T;\\
		\alpha^{14} & = & 1+ & & & \alpha^3& = (1,0,0,1)^T;\\
		\end{array}$
	\end{center}
	
	Consider the Goppa Code $\Gamma(L,g(z))$ defined by $$g(z)=(z+\alpha)(z+\alpha^{14})=z^2 + \alpha^7z+1,$$ $$ L=\{\alpha^i \, | \, 2 \leq i \leq 13\}.$$
	
	Now, in order to find the Parity check matrix of $ \Gamma(L,g(z))$, we need to compute
	$g(\alpha^2)^{-1}=(\alpha^4 +\alpha^9 +1)^{-1}=((0,0,0,1)^T)^{-1}=\alpha^{12}$, 
	and similarly, other entries to compute $H$ as described in Remark \ref{parmat}.
	$$H= \left( \begin{array}{cccccccccccc} 
	\alpha^9 & \alpha^{10} & \alpha^9 & \alpha^{14} & \alpha^6 & 0 & \alpha^{10} & \alpha^8 & \alpha^2 & \alpha^7 & \alpha^{14} & \alpha^6 \\
	\alpha^{12} & \alpha^6 & \alpha^6 & \alpha & \alpha^{11} & 1 & \alpha^{14} &  \alpha^8 & \alpha^{11} & \alpha^{14} & \alpha^{12} & \alpha \end{array} \right).$$
	The entries can be observed as binary vectors of length $4$ using the table described above. This is given by:
	$$H= \left( \begin{array}{cccccccccccc} 
	0&1&0&1&0&0&1&1&0&1&1&0\\
	1&1&1&0&0&0&1&0&0&1&0&0\\
	0&1&0&0&1&0&1&1&1&0&0&1\\
	1&0&1&1&1&0&0&0&0&1&1&1\\
	1&0&0&0&0&1&1&1&0&1&1&0\\
	1&0&0&1&1&0&0&0&1&0&1&1\\
	1&1&1&0&1&0&0&1&1&0&1&0\\
	1&1&1&0&1&0&1&0&1&1&1&0\\ \end{array} \right).$$
	
	Then, the Null-space of $H$ produces generator matrix of $\Gamma(L,g(z))$. Hence the generator matrix $G$ becomes:
	
	$$G= \left( \begin{array}{cccccccccccc} 
	1&1&1&1&0&1&0&1&0&1&0&0\\
	0&1&0&0&1&1&1&1&0&0&1&0\\
	0&0&1&0&1&0&1&1&1&0&0&0\\
	0&1&0&1&0&0&1&1&0&0&0&1\\ \end{array} \right).$$
	
\end{example}  $\hfill \Box$

%
%

\subsection{Encoding with Goppa Codes}

Let $\Gamma(L,g(z))$ be a Goppa code, where $g(z)$ is some primitive polynomial with $deg(g(z))=t$ and $|L|=n$. Let dim$_{\F_q}(\Gamma(L,g(z)))=k$, and $G$ be `$k \times n$' sized generator matrix for respective Goppa code; then encoding of a $k$-length message vector $\textbf{m}$ over $\F_q$ is $\textbf{m} G$.

\subsection{Correction of errors/ Syndrome decoding of Goppa codes}

Let the vector $\textbf{y} = (y_1, y_2,\dots, y_n)$ be received with `{$r$}' number of errors, where `$2r+1 \leq d$' (for maximum number of error correction). 
Let $L=\{\alpha_1,\alpha_2,\dots,\alpha_n\}$, 
$$\textbf{y}=(y_1, y_2,\dots, y_n)= \underbrace{(c_1,c_2,\dots,c_n)}_{\text{codeword}}+\underbrace{(e_1,e_2,\dots,e_n)}_{\text{error vector}}$$ with $e_i \neq 0$ at exactly $r$-places. We need to 
\begin{itemize}
	\item locate positions of error (say $B=\{i: 1\leq i \leq n \text{ and } e_i \neq 0\}$);
	\item find the corresponding error values (values of $e_i : i \in B$).
\end{itemize}

In order to find these, we define two polynomials

\begin{dfn}
	Error locater polynomial $\sigma(z)$ and Error evaluator polynomial $w(z)$
	\begin{itemize}
		\item $\sigma(z):= \prod_{B} (z-\alpha_i)$ \space (this is a `r' degree polynomial);
		\item $w(z):=\sum_{i \in B} e_i \prod_{j \in B; j\neq i} (z- \alpha_j)$ \space (this is a `$r-1$' degree polynomial). 
	\end{itemize}
\end{dfn}

\begin{dfn}
	Syndrome of received vector $\textbf{y}$ is defined as $S(\textbf{y})$ where: 
	
	$$\begin{array}{ll}
	S(\textbf{y})&:= \sum_{i=1}^n \frac{y_i}{z-\alpha_i}  \vspace{2mm} \\
	& =\sum_{i=1}^n \frac{c_i}{z-\alpha_i} + \sum_{i\in B} \frac{e_i}{z-\alpha_i} \vspace{2mm}  \\
	& =\sum_{i\in B} \frac{e_i}{z-\alpha_i}  \mod (g(z)).
	\end{array}$$
\end{dfn}

\begin{proposition}
	Let $\textbf{e}$ be the error vector having weight $r: r \leq \left\lfloor \frac{t}{2} \right\rfloor$. Let $\sigma(z), w(z)$ and $S(\textbf{y})$ be as described above. Then the following properties hold:
	\begin{enumerate}[$(i)$]
		\item deg$(\sigma(z)) =r$;
		\item deg$(w(z))\leq r-1$;
		\item gcd$(\sigma(z),w(z))=1$;
		\item $e_k=w(\alpha_k)/\sigma'(\alpha_k)$, where $k \in B$ and $\sigma'$ represents derivative of $\sigma$;
		\item $\sigma(z)S(\textbf{y}) \equiv w(z)$  (mod $g(z)$).
	\end{enumerate}
\end{proposition}

\textbf{Error-correction:} \underline{Algorithm for correcting $r\leq \left\lfloor \frac{t}{2} \right\rfloor$ errors in a Goppa code:}\\
\begin{enumerate}[Step ($i$):]
	\item Compute the syndrome $$S(\textbf{y})=\sum_{i=1}^n \frac{y_i}{z-\alpha_i},$$
	\item Solve the key equation $$\sigma(z)S(\textbf{y}) \equiv w(z) \text{ (mod g(z))},$$ by writing $$\sigma(z)=\sigma_0 + \sigma_1 z + \cdots + \sigma_{r-1}z^{r-1}+z^r,$$ $$w(z)=w_0 + w_1 z + \cdots + w_{r-1}z^{r-1},$$ and solve for $t$ equations and $2r$ unknowns. \\If the code is binary, take $w(z)=\sigma ' (z)$,
	\item Determine the set of error locations $B=\{i \,\,:\,\, 1 \leq i \leq n \text{ and } \sigma(\alpha_i)=0\}$,
	\item Compute the error values $e_i = \frac{w(\alpha_i)}{\sigma ' (\alpha_i)}$ for all $i \in B$,
	\item The error vector $\textbf{e}=(e_1,e_2,\dots,e_n)$ is defined by $e_i$ for $i \in B$ and zeros elsewhere,
	\item The codeword sent is calculated as $\textbf{c} =\textbf{y} -\textbf{e}$.
\end{enumerate}

\subsection{Patterson's Algorithm for Error Correction}
The patterson algorithm decodes only binary Goppa codes. It computes the syndrome $S(\textbf{y})$ of a received vector and then solves the key equation 
$\sigma(z)S(\textbf{y}) \equiv w(z) \text{ (mod g(z))}$
with $w(z) = \sigma '(z)$ by heavily exploiting the requirement that the code is binary. 
The error locater polynomial can be split in even and odd powers of $z$ such that $\sigma(z) = a^2(z) + z b^2(z)$, as field has characteristic $2$.

The Patterson algorithm can be described as below:
\medskip
\vspace{1mm}
\hrule
\vspace{1mm}\textbf{Input:} The received vector \textbf{y} and the Goppa code $\Gamma(L,g)$.
\vspace{0.5mm}
\hrule 
\begin{enumerate}[Step $(i)$:]
	\item Compute syndrome $S(\textbf{y})$ an element of $\frac{\F_{q^m}[z]}{\left<g(z)\right>}$
	\item Compute $T(z) = S(\textbf{y})^{-1} \mod g(z)$
	\item Compute $P(z) = \sqrt{T(z)+z} \mod g(z)$
	\item Compute $u(z)$ and $v(z)$ with $u(z)=v(z)S(\textbf{y}) \mod g(z)$
	\item Compute the locater polynomial $\sigma(z) = u(z)^2 + z v(z)^2$
	\item Find the roots of $\sigma(z)$
	\item Find error positions, i.e., error vector \textbf{e}\vspace{1mm}
\end{enumerate}
\vspace{-2mm}
\hrule
\vspace{1mm}\textbf{Output:} The error vector \textbf{e}.
\vspace{0.5mm}
\hrule

\subsection{Decoding the Message after Discovering the Codeword}

After correcting possible errors in a codeword, one can find the message sent by recalling that $$(m_1, m_2, \dots, m_k) \cdot G = (c_1,c_2,\dots, c_n),$$ equivalently,
$$ G^T \cdot \left(\begin{matrix} m_1 \\\vdots\\ m_k	\end{matrix} \right) = \left( \begin{matrix} c_1 \\ \vdots \\ c_n \end{matrix}\right),$$

to find the message vector $(m_1,m_2,\dots,m_k)$, one reduces above system to

\begin{center}
	
	$\left[ \begin{tabular}{c|c}
	& $c_1$\\ 
	$G^T$ & $\vdots$ \\
	& $c_n$ \\
	\end{tabular} \right] 
	\sim 
	\dots 
	\sim 
	\left[ \begin{matrix}
	{\begin{tabular}{ccc|c}
		&	  &	   & $m_1$\\
		&$I_k$&    & $\vdots$ \\
		&	  &	   & $m_k$ 
		\end{tabular}}\\
	\hline 	  \\
	P\\
	\\
	\end{matrix} \right]$
	
\end{center}


\begin{example}
	Let $\F_{3^2}$ be the field corresponding to the primitive polynomial $x^2 -x-1$ over the base field $\F_3$ and let `$\alpha$' be one of its root. Then we have $$ \begin{array}{rrrrr}
	0=& & &=&(0,0)^T;\\
	1=&1& &=&(1,0)^T;\\
	\alpha=& &\alpha&=&(0,1)^T;\\
	\alpha^2=&1 &+\alpha&=&(1,1)^T;\\
	\alpha^3=&1 &-\alpha&=&(1,-1)^T;\\
	\alpha^4=&-1& 		&=&(-1,0)^T;\\
	\alpha^5=&  &-\alpha&=&(0,-1)^T;\\
	\alpha^6=&-1&-\alpha&=&(-1,-1)^T;\\
	\alpha^7=&-1&+\alpha&=&(-1,1)^T;\\
	\end{array}$$
	Consider the Goppa code $\Gamma(L,g(z))$ defined by $$ \begin{array}{c}
	g(z)=z(z-\alpha^7)=z^2+\alpha^3z,\\ L=\{\alpha^i \, : \, 0 \leq i \leq 6\}. \end{array}$$
	Then the Parity check matrix over $\F_3(\alpha)$ will be $$H= \left( \begin{array}{ccccccc}
	\alpha^4&\alpha^3&\alpha^2&\alpha&1&\alpha^7&\alpha^6\\
	\alpha^6&\alpha^3&\alpha^6&\alpha^2&\alpha^3&\alpha^5&\alpha^5
	\end{array}  \right);$$
	Equivalently, the Parity check matrix over $\F_3$ will be $$H= \left( \begin{array}{rrrrrrr}
	-1&1&1&0&1&-1&-1\\
	0&-1&1&1&0&1&-1\\
	-1&1&-1&1&1&0&0\\
	-1&-1&-1&1&-1&-1&-1\\ \end{array} \right).$$
	This gives the generator matrix as $$G= \left( \begin{array}{rrrrrrr}
	-1&0&-1&1&0&0&0\\
	-1&0&-1&0&1&1&0\\
	-1&1&-1&0&0&0&1 \end{array}\right).$$ 
	The parameters of this code are $[7,3, \geq3]$.
	Now, let the message $\textbf{m}=(0,0,0)$ be sent. Firstly, encoding of this vector will be $\textbf{c}=\textbf{m}G=(0,0,0,0,0,0,0)$. Suppose that the vector $\textbf{y}=(0,0,0,0,0,0,-1)$ is received having one error. Our aim is to find the error vector.
	\begin{enumerate}[$(i)$]
		
		\item Syndrome \vspace{-1mm}$$S(\textbf{y})= \sum_{i=1}^6 \frac{y_i}{z-\alpha_i}= \frac{-1}{z-\alpha^6} \equiv \alpha^2 +\alpha z \mod g(z),$$\vspace{-1mm}
		
		\item Substituting $\sigma(z) = \sigma_0 + z$ and then computing $\sigma(z)S(\textbf{y})$ $\mod z^2 + \alpha^3 z$ gives 	
		\vspace{-1mm}
		$$\begin{array}{rl}
		\sigma(z)S(\textbf{y})=& (\sigma_0+z)(\alpha^2+\alpha z)\\
		=&\sigma_0\alpha^2 + (\alpha^2+ \alpha \sigma_0)z + \alpha z^2 \\
		\equiv& \alpha^2 \sigma_0 + (\alpha^2 +\alpha \sigma_0 \alpha^4)z\\
		=& \alpha^2 \sigma_0 +(\alpha^7 +\alpha \sigma_0)z 
		\end{array}$$\vspace{-1mm}
		
		Thus, for $w(z) = w_0$, we get the system of equations by comparing coefficients of $\sigma(z) S(\textbf{y}) \equiv w(z) \mod g(z)$ as:
		$$ \left\{ \begin{array}{cl} 
		w_0& = \alpha^2 \sigma_0,\\ 
		0&=\alpha^7+\alpha \sigma_0. 
		\end{array} \right. $$ 
		The solution to this system is $\sigma_0=\alpha^2, w_0=\alpha^4$ and hence $\sigma(z)=z+\alpha^2$ and $w(z)=\alpha^4$.
		
		\item The root of $\sigma(z)$ is $\alpha^6=\alpha^7$, thus the set of error locations is $$B=\{i \, | \, \sigma(\alpha_i)=0\}=\{7\}.$$\vspace{-8mm}
		
		\item The error value $e_7=\frac{\alpha^4}{1}=\alpha^4=-1$.
		
		\item The codeword sent must have been \vspace{-1mm}$$\textbf{c}=\textbf{y}-\textbf{e}=(0,0,0,0,0,0,0).$$
	\end{enumerate}
	Then the original message can be found out solving the augmented system $\left[G^T \, \bigg| \, \textbf{c}^T\right]$, which results $(0,0,0).$
\end{example}

With this, we now proceed to cryptographic applications of coding theory in the coming sections.

\section{Code-based Cryptography}

The birth of code-based cryptography was inspired by the work of Robert J. McEliece in 1978.
He was the first one to implement the use of binary Goppa codes to develop code-based public key cryptosystem.
There are several reasons why Goppa codes are the primary choice for the McEliece cryptosystem. 
First of all, Goppa codes have a fast polynomial time decoding algorithm. Another reason is that Goppa codes are ``easy to generate but hard to find''. 
Any irreducible polynomial over a finite field $\F_{2^m}$ can be used to create a Goppa code, but the generator matrices of Goppa codes are nearly random. 

For any fixed length $n$, there are many different Goppa codes. 
Though the exact number of Goppa codes, given length $n$ of the code and degree $t$ of the Goppa polynomial, is not known, Ryan and Fitzpatrick \cite{PFitzpatrick} found a way to calculate upper bounds, which are exact for some of the small parameters. For example, the upper bound for the number of Goppa codes of length $128$ which are able to correct at least $10$ errors is $1037499670492467 \approx 1.04 \times 10^{15}$, while the upper bound for Goppa codes of the same length able to correct at least $15$ errors is $23765478069520611201643781 \approx 2.38 \times 10^{25}$. In fact, the number of Goppa codes grows exponentially with the length of the code and the degree of the generating polynomial \cite{PFitzpatrick}. Goppa codes are still the primary family of codes used with the McEliece cryptosystem.

Following to this, Niederreiter used parity check matrix of the Generalized Reed-Solomon codes as public key to develop public key cryptosystem.
Pursuing the same way, came use of Reed Solomon, BCH codes, Reed-Muller codes, all of whose security reduced to standard hard problem of coding theory, which we will describe in this section. Majority of these variants were broken or lack the security proof.
Hamming Quasi Cycic (QC), Quasi-Cyclic Moderate density parity-check (QC-MDPC), Rank mertic, LRPC codes based schemes also form code-based cryptosystems. 

\subsection{Cryptosystems}

Essentially, there are two types of code-based cryptosystems, upon whose structure dwells all other cryptosystems in this class.
The first system is the McEliece Cryptosystem, in which the generating matrix of the Goppa code is hidden by scrambling and permuting the entries of that matrix, and making it public. 
The ciphertext is generated by encoding message with the matrix available in public key and x-oring with some small weight error, depending on parameters of Goppa code. 
The second system is the Niederreiter cryptosystem, in which message is random small weighted error vector. 
The public key becomes the scrambled-permuted parity check matrix of Generalized Reed Solomon (GRS) code. Niederreiter's proposed GRS codes were shown to be a bad choice in his cryptosystem by Sidelnikov, Shestakov \cite{VMSidelnikov}, but Goppa codes were found to be working fine.

Since a large public key size is one of the drawbacks of code-based cryptosystems, there have been many proposals attempting to reduce the key size.
Examples of this include Quasi-cyclic, (QC) as well as low density parity check (QC-LDPC) codes.
Recently, there have been several publications on structural attacks against such highly structured codes.
Otmani et al. \cite{otmani} cryptanalysed a McEliece cryptosystem based on QC-LDPC codes. The attack exploits the QC structure to find some version of the secrret key, then uses Stern's algorithm to reconstruct the entire secret key.
Faigére et al. presented an algebraic attack against the McEliece cryptosystem using non-binary QC codes at Eurocrypt 2010. The attacker sets up a system of algebraic equations, the solution of which will be an alternant decoder for the underlying code.
While this system can't be solved efficiently for the original McEliece cryptosystem, the additional QC structure allows to significantly reduce the number of unknowns of this system.

In 2010, Faugére et al. presented a Goppa code distinguisher, which allows to distinguish a goppa code from random codes, provided the code rate is very high (code rate is dimension of code divided by the code length).
This is useful for security proof of such a cryptosystem.

\subsection{Hard Problems in Coding Theory}

The general problems of coding theory which describes the security behind code-based cryptosystems are listed below.

\begin{enumerate}[\textbf{Problem }$1.$]
	\item \textbf{General Decoding Problem:} Given an $[n,k]$ code $\mc{C}$ over $\F_q$, an integer $t_0$ and a vector $\textbf{c} \in \F_q^n$, find a codeword $\textbf{x} \in \mc{C}$ with $d(\textbf{x}, \textbf{c}) \leq t_0$.
	
	\item \textbf{Syndrome Decoding (SD) Problem:} Given a matrix $H$ and a vector $\textbf{s}$, both over $\F_q$, and a non-negative integer $t_0$; find a vector $\textbf{x} \in \F_q^n$ with Hamming weight $wt(\textbf{x})= t_0$ such that $H \textbf{x}^T =\textbf{s}^T$.
	
	These problems were proved to be NP-complete in $1978$ by Berlekamp et al. \cite{EBerlekamp} for binary codes and in 1997 by Alexander Barg \cite{Alexander} for codes over all finite fields.

	\item  \textbf{Goppa Parameterized Syndrome Decoding (GPSD):} Given a binary matrix $H$ of size $2^m \times r$ and a syndrome $\textbf{s}$, decide whether there exists a codeword $\textbf{x}$ of weight $r/m$ such that $H \textbf{x}^T = \textbf{s}^T$.
	
	This problem is also an NP-complete problem, proof of which was given by Finiasz \cite{Finiasz}
	
	\item \textbf{Goppa Code Distinguishing (GD):} Given an $r\times n$  matrix $H$, decide whether $H$ is the parity check matrix of a Goppa code.
	
	In 2013, Faugére - Gauthier - Umaña - Otmani - Perret - Tillich \cite{FGUOPT} showed that ``high rate'' binary Goppa codes can be distinguished from random linear codes. However it does not work at
	\begin{itemize}
		\item $8$ errors for $n =1024$ (where McEliece used 50 errors)
		\item $20$ errors for $n = 8192$ (a variant of classic mceliece).
	\end{itemize}
	
\end{enumerate}

\subsection{Information-Set Decoding}

An attacker who got hold of an encrypted message \textbf{y} has two possibilities in order to retrieve the original message \textbf{m}.

\begin{itemize}
	\item Find out the secret code; i.e., find the generating matrix $G$ given public key $\hat{G}$ which is scrambled-permuted generating matrix, or
	
	\item Decode $\textbf{y}$ without knowing an efficient decoding algorithm for the public code given by $\hat{G}$.
\end{itemize}

Attacks of the first type are called \textit{structural attacks}. If $G$ or an equivalently efficiently decodable representation of the underlying code can be retrieved in sub-exponential time, this code should not be used in the McEliece cryptosystem. 
Suitable codes are such that the best known attacks are decoding random codes. We will describe how to correct errors in a random-looking code with no obvious structure

\begin{dfn}[Information Set]
	Let $G$ be a generator matrix of a $[n,k]-$linear code, $I$ be a subset of $\{1,\dots,n\}$ and $G_I$ be the $k \times k$ sub-matrix of $G$ defined by the columns of $G$ with indices from $I$. If $G_I$ is invertible then $I$ is an information set. 
\end{dfn}

An equivalent definition from parity check matrix point of view:
Using a parity-check matrix $H$, an information set $I$ implies the non-singularity of the sub-matrix formed by the columns with indices $\{1,2,\dots,n\} \setminus I$. The description in terms of parity-check matrices, although less intuitive, favors the explanation on how ISD algorithms work. \\

Information-set decoding (ISD) induces a generic attack against all code based cryptosystems regardless of our current scheme. 
The basic ISD algorithm was given by Prange \cite{prange} with improvements by Leon \cite{leon}, Lee-Brickell \cite{lee}, Stern \cite{stern} and Canteaut-Chabaud \cite{canteaut}.

An attacker does not know the secret code and thus has to decode a random-looking code without any obvious structure.
The best known algorithms which do not exploit any code structure rely on information-set decoding, an approach introduced by Prange. 
The idea is to find a set of coordinates of a garbled vector which are error-free (i.e., an Information-Set, as defined above) and such that the restriction of the code's generator matrix to these positions is invertible. 
Then, the original message can be computed by multiplying the encrypted vector by the inverse of the submatrix.
\bigskip

\section{McEliece Cryptosystem}

Recent public-key cryptography is largely based on number theory problems, such as factoring or computing discrete logarithm. These systems constitute an excellent choice in many applications, and their security is well defined and understood. One of the major drawbacks, though, is that they will be vulnerable once quantum computers of an appropriate size are available. There is then a strong need for alternative systems that would resist attackers equipped with quantum technology.

With the development of Quantum Computers, the risk to present day cryptography is increasing. The coming scenario to cryptographic world relies upon Post-Quantum Cryptosystems, or we can say Quantum resistant cryptosystems. 
Coding Theory based encryption systems are one kind of cryptosystems that are able to resist quantum computing, and this provides an area in Post-Quantum Cryptography.

\textbf{Robert J. McEliece} (born $1942$) is a mathematician and engineering professor at Caltech. He was the $2004$ recipient of \textit{Claude E. Shanon Award} and the $2009$ recipient of the \textit{IEEE Alexander Graham Bell Medal}. He gave the notion of code-based cryptography and developed the public key cryptosystem based on binary Goppa codes  in $1978$, namely \textit{McEliece} cryptosystem.

The general idea of security behind this cryptosystem is the hardness in decoding a random linear code (Problem 1). 
Except for the choice of parameters, this cryptosystem is unbroken till now. 
While the huge size of key remains an issue, yet it is fair enough to prove efficient encryption system.
This system did not get that esteem which other cryptosystems of that time have got due to handling of keys issue.
When Shor's algorithm appeared to impact number theory based cryptosystems, the value of code-based cryptosystems ascended and McEliece cryptosystem being the oldest one in that cluster got the significant research.

In this section we describe the McEliece cryptosystem in detail, covering its weaknesses and applications. To be accurate, this cryptosystem is a code-based system and the underlying code is the famous Goppa code.
In the following sections, we fix the parameters of Goppa code as:
\begin{center}
	\begin{tabular}{lcl}
		$n$ &: &length of the code;\\
		$k$ &: &the dimension of code over the field $\F_q$;\\
		$t$ &: &the degree of Goppa polynomial.
	\end{tabular}
\end{center}
\vspace{2mm}
The original version of McEliece cryptosystem given by Robert J. McEliece\cite{rjm}, based on binary Goppa codes in the year $1978$ is described as follows. The values of $n, k$ and $t$ are publicly available parameters, but $L, g, P$ and $ S$ are randomly generated secrets. Then, this public-key cryptosystem work as follows:

\begin{enumerate}[Step 1:]
	
	\item Firstly \textit{Alice} generates a public and private key pair depending upon publicly available values. During this,
	
	\begin{enumerate}[$(i)$]
		
		\item Alice selects a binary {$[n,k]$-Goppa code}, with its `{$k\times n$}' sized generator matrix {$G$}, capable of correcting `{$t$}' errors;
		
		\item She then selects a random `{$k \times k$}' binary non-singular matrix {$S$} and a `{$n \times n$}' permutation matrix {$P$};
		
		\item She computes the `{$k \times n$}' matrix {$\hat{G}= S \cdot G \cdot P$};
		
		\item She publishes her public key : {$\left( \hat{G}, t \right)$};
		
		\item She keeps her private key : {$(S, G, P)$}.
		
	\end{enumerate}

	\item Suppose \textit{Bob} has to send an encrypted message to \textit{Alice}:
	
	\begin{enumerate}[$(i)$]
		
		\item Bob has a binary plaintext-message {$\textbf{m}$} of length `{$k$}';
		
		\item He loads the public key of Alice : {$\left( \hat{G}, t \right)$};
		
		\item He generates a random {$n$}-bit vector {$\textbf{z}$} with Hamming weight `{$t$}';
		
		\item Bob computes the ciphertext {$\textbf{c} = \textbf{m} \cdot \hat{G} + \textbf{z}$} and sends to Alice.
		
	\end{enumerate}
	
	\item Suppose \textit{Alice} have received the ciphertext \textbf{c}. She decrypts the received ciphertext as:
	
	\begin{enumerate}[$(i)$]
		
		\item Alice computes {$P^{-1}$} using her private key;
		
		\item Post multiplication by $P^{-1}$, she computes {$\textbf{c} \cdot P^{-1} = \textbf{m} \cdot S \cdot G + \underbrace{\textbf{z} \cdot P^{-1}}_{\text{this has weight t}}$};
		
		\item Finally, she uses the decoding algorithm (Patterson's algorithm) of Goppa codes for the secret Goppa code to determine the value of {$\textbf{m}$}.
		
	\end{enumerate}
	
\end{enumerate}

There are a number of decoding algorithms for Goppa codes. In usual, Patterson's decoding algorithm is followed as it make use of binary irreducible Goppa codes

\subsection{Information-Set Decoding Attack}

At PQCrypto 2008, several speed ups for ISD techniques were proposed by Bernstein et al.\cite{DJBLength} which led improvements in reducing the cost to attack the original McEliece parameters $(1024, 524, 50)$ to $2^{60.5}$ binary operations.
Finiasz and Sendrier \cite{Finiasz} presented a further improvement which could be combined with the improvements in \cite{DJBLength} but did not analyse the combined attack.
It was proved in \cite{DJBLength} that to obtain a $128$-bit security, the Goppa codes must have length $2960$ and dimension $2288$ with a degree-56 Goppa polynomial and 57 added errors. 

Let $\hat{G}$ be the public key of McEliece cryptosystem.
Then, for a message \textbf{m}, ciphertext \textbf{c} is obtained as 
$ \textbf{m} \cdot \hat{G} + \textbf{e}$; equivalently, we have:
\begin{center}
	\begin{tabular}{lll}
		$\textbf{m}\cdot \hat{G} + \textbf{e}$ &$=$& $\textbf{m}_{1 \times k} \cdot \left( G_1, G_2, \cdots, G_n \right)_{k \times n} + (e_1,e_2, \cdots, e_n)_{1 \times n}$\\
		&$=$& $\left( \textbf{m}G_1, \textbf{m}G_2, \cdots, \textbf{m}G_n \right)+(e_1,e_2, \cdots, e_n)$\\
		&$=$& $\left( \textbf{m}G_1+e_1, \textbf{m}G_2+e_2, \cdots, \textbf{m}G_n+e_n \right)$
	\end{tabular}\\
\end{center}
where, for $1 \leq i \leq n, G_i$ represents $i^{th}-$column of the Scrambled-Permuted generated matrix of the code which is the given public key matrix. 

Here, a critical point is that the Hamming weight of error vector $wt(\textbf{e})=t$ which is very small as compared to the block length of code. 
This means only $t$ out of $n$ coordinates of $\textbf{e}$ are non-zero.
Apparently, if a cryptanalyst could guess `$k$' out of `$n-t$' coordinates from \textbf{c} that corresponds to `$0$' at that coordinate of \textbf{e}, then the restriction to those `$k$' columns of \textbf{c} and the Public key $\hat{G}$ is observed as: $ \overline{\textbf{c}} = \textbf{m} \cdot \overline{\hat{G}}$.
For such an instance, suppose $\{i_1,i_2,\dots, i_k\} \subset \{ 1,2, \dots,n \}$ be such that for each $1 \leq j \leq k, e_{i_j} =0$.
Then, upon considering the restriction of public key on these indices we arrive at:
\begin{center}
	\begin{tabular}{lll}
		$\underbrace{(c_{i_1},c_{i_2},\dots,c_{i_k})_{1 \times k}}_{\overline{\textbf{c}}}$ &$=$& $\textbf{m}_{1 \times k} \cdot \underbrace{\left(G_{i_1}, G_{i_2}, \cdots, G_{i_k} \right)_{k \times k}}_{\overline{\hat{G}}}$
	\end{tabular}
\end{center}

This means, if the `$k \times k$' sized matrix $\overline{\hat{G}}$  is invertible, then the message \textbf{m} can be recovered by just post multiplying by inverse of $\overline{\hat{G}}$.
It comes out to be that it requires $ {n \choose k}\bigg/{{n-t} \choose k}$ number of guesses to succeed and the work factor comes to be 
$$k^3 \cdot \frac{{n \choose k}}{{{n-t} \choose k}} \approx k^3\left( 1 - \frac{t}{n}\right)^{-k} $$
where $k^3$ is cost of inverting a $k \times k$ matrix. 
The original parameters proposed in McEliece cryptosystem are:

\begin{center}
	\begin{tabular}{lll}
		Length of the code $n$&:& 1024,\\
		The binary extension field $\F_{2^m}$&:&$\F_{2^{10}}$ i.e., $m=10$,\\
		The degree $t$ of Goppa polynomial&:&50,\\
		The dimension $k$ of the Goppa code&:&$k=n-mt=524$.  
	\end{tabular}
\end{center}

For original parameters of McEliece cryptosystem, the work factor to find an information set comes out to be:  $$ \sim 1.9 \times 10^{24} \simeq 2^{79.7}.$$
Hence it is not an appropriate algorithm to perform such an attack.

Looking from another perspective that the public key $\hat{G}$ is again a generator matrix for some code with minimum distance at least $2t+1$. We consider two cases for two messages \textbf{u} and $\textbf{u}'$.

\textit{Case} $(i)$: If $\textbf{u} \neq \textbf{u}'$, then $d(\textbf{u}\hat{G}, \textbf{u}'\hat{G}) > 2t$, i.e., $wt_H(\textbf{u}\hat{G}+\textbf{u}\hat{G})>2t$. Now if $wt(\textbf{e}) =t$, we have $wt_H(\textbf{u}'\hat{G}+\textbf{u}\hat{G}+ \textbf{e})>t$.

\textit{Case} $(ii)$: If $\textbf{u} = \textbf{u}'$, then $\textbf{u}\hat{G}= \textbf{u}'\hat{G}$. Now if $wt(\textbf{e}) =t$, we have $wt_H(\textbf{u}'\hat{G}+\textbf{u}\hat{G}+ \textbf{e})=t$.

From this, Eavesdropper upon receiving a ciphertext $\textbf{c}=\textbf{u} \hat{G} + \textbf{e}$, guesses a message $\textbf{u}'$ and checks $wt(\textbf{u}'\hat{G} + \textbf{c})$. It this is not equal to $t$, then he makes sure that $\textbf{u} \neq \textbf{u}'$. If the error vector $\textbf{e}$ was chosen in such a way that $wt(\textbf{e})\leq t$. Then also the similar arguments work providing eavesdropper now checks $wt(\textbf{u}'\hat{G} + \textbf{c}) \leq t$.
There are many improvements for this attacks, and a few algorithms which execute this attack more effectively: Lee-Brickell's algorithm and Stern's algorithm.\\


The \textbf{Lee-Brickell} algorithm to recover error vector \textbf{e} from original McEliece cryptosystem using Information set decoding is explained as:\\

\begin{tabular}{lll}
	
	\texttt{Input} &:& A generator matrix $G$, a ciphertext \textbf{y}$\in \F_q^n$ and a parameter $p \in \mathbb{N}$.\\
	
	\texttt{Output} &:& An error vector \textbf{e} of weight $t$.\\
	
\end{tabular}
\hrule
\begin{enumerate}[Step $(i)$]
	\vspace{-4mm}\item Choose a random information set $I$ of size $k$ and compute $\textbf{y}_I, G_I$ choosing corresponding columns of $G$ and $G'= G_I^{-1} G$ if inverse exists.
	\vspace{-3mm}
	\item Calculate $\textbf{y}' = \textbf{y} - \textbf{y}_I G_I'$.
	\vspace{-2mm}
	\item For each size-$p$ subset $\{a_1,\dots,a_p\} \subset I$, for each $x_1, x_2, \dots, x_p \in \F_q \setminus \{0\}$, compute the vector $\hat{\textbf{g}}= \sum_{i=1}^p x_i G_{a_i}'$
	\vspace{-1mm}
	\item Set $\textbf{e} = \textbf{y}' - \hat{\textbf{g}}$. If $wt(\textbf{e})=t$ then return \textbf{e}.
	\vspace{-1mm}
	\item Go back to step $(i)$. 
	\vspace{-4mm}
\end{enumerate}		
\hrule
\vspace{2mm}

We indicate with $G_j'$ the row of $G'$ where there is a $1$ in position $j$. Note that, by definition, this is unique if $j$ is an element of an information set.

The minimum work factor for this implementation comes out to be $\simeq 2^{73.4}$ for original parameters set.
In \cite{peters}, Peters generalised Stern's and Lee-Brickell's algorithms (both are variants of ISD attacks) on $\F_q$. 

Subsequently, we have generalized Stern's algorithm, ball collision decoding algorithm by Bernstein, Lange and Peters, Sendrier's Decoding One Out of Many algorithm, and many more. 
It is to be noted that this is a per-message attack; the secret key of the system still remains unknown to the cryptanalyst.
The last improvement of this algorithm refers to Kruk \cite{krouk}, who proposed a solution to reduce its complexity, thus obtaining a work factor equal to $2^{59}$ for the original parameters.

The above process is demonstrated in MATLAB and is achieved with help of a few examples. The code can be referred from Appendix.

\subsection{Message-Resend or Related message attack}

Suppose that the sender encrypted a message \textbf{m} twice and two ciphertexts are generated 
$$\left\{
\begin{matrix}
\textbf{c}_1 &= &\textbf{m} \cdot S \cdot G \cdot P + \textbf{e}_1\\
\textbf{c}_2 &= &\textbf{m} \cdot S \cdot G \cdot P + \textbf{e}_2\\
\end{matrix}
\right.$$
where $\textbf{e}_1 \neq \textbf{e}_2$.
This is called message-resend condition. In this case it is easy for the cryptanalyst to recover \textbf{m} from the above system. As same message is encrypted twice, we say resend depth is $2$ in this case. 
%
Let $\textbf{c}_j(i)$ be $i^{th}$ coordinate of $\textbf{c}_j$, then
\vspace{-2mm}
\begin{center}
	$L_0 := \{i \in \{1,2,\dots,n\} : \textbf{c}_1(i) + \textbf{c}_2(i) = \textbf{e}_1(i) + \textbf{e}_2(i) = 0\}$;\\
	$L_1 := \{i \in \{1,2,\dots,n\} : \textbf{c}_1(i) + \textbf{c}_2(i) = \textbf{e}_1(i) + \textbf{e}_2(i) = 1\}$.
\end{center}

\begin{itemize}
	
	\item $l \in L_0$ means either $\textbf{e}_1(l) = 0 = \textbf{e}_2(l)$ or $\textbf{e}_1(l) = 1 = \textbf{e}_2(l)$. Assuming the event of choosing error vectors are independent, we have
	$$Pr \left( \textbf{e}_1(l) = 1 = \textbf{e}_2(l) \right) = \left( \frac{t}{n} \right)^2.$$
	For the case of original parameters of McEliece cryptosystem, it is $(50/1024)^2 \approx 0.0024$. So, when we consider $l \in L_0$, most significant is the case when $\textbf{e}_1(l) = 0 = \textbf{e}_2(l)$; equivalently, neither $\textbf{c}_1(l)$ nor $\textbf{c}_2(l)$ is garbled by error vectors.
	
	\item $l \in L_1$ certainly means one of $\textbf{c}_1(l)$ or $\textbf{c}_2(l)$ is garbled by error vector.
	
\end{itemize}

Now our aim is to approximate the probability of guessing $k$ ungarbled columns from those indexed by $L_0$. 
Let $p_m$ be the probability that precisely $m$ coordinates are garbled by $\textbf{e}_1$ and $\textbf{e}_2$. Then 
$$p_m = Pr\left( |\{i: \textbf{e}_1(i) =1\} \cap \{i : \textbf{e}_2(i) =1\}|= i \right) = \frac{{t \choose i} {n-k\choose t-i}}{{n \choose t}}$$

Therefore, the expected cardinality of $L_1$ is $$E(|L_1|) = \sum_{m=0}^{t} (2t - 2m)p_m$$ since every $i$ for which $\textbf{e}_1(i)=1=\textbf{e}_2(i)$ reduces $|L_1|$ by two.

For McEliece cryptosystem's original parameters set, this comes out to be  $\approx 95.1$.

For example, suppose $|L_1|= 94$. Then $|L_0|= 1024-94 =930$, of which $|L_0| \times 0.0024 \approx 3$ are garbled. 
We have the probability of guessing $5413$ ungarbled columns from those indexed by $L_0$ is 
$$\frac{{927 \choose 524}}{{930 \choose 524}} \approx 0.0828.$$
So the cryptanalyst expects to succeed in this case with only $12$ guesses, at a cost of $12 \times 524^3 \approx  10^{10}$. 	
These results are a factor of $10^{15}$ better than exhaustive information-set decoding attack as analyzed above.

The conclusions of the literature are that information set decoding is an efficient method of attacking the McEliece system but that from a practical viewpoint the system is unbreakable provided the code is long enough. Bernstein et al. \cite{DJBLength} give recommended code lengths and their corresponding security, as described below. 

\begin{center}
	\begin{tabular}{|c|c|c|}
		\hline
		\textcolor{blue}{Length $n$ of code} 	&	\textcolor{blue}{Weight $t$ of error vector}	&	\textcolor{blue}{Security (in bits)} \\
		\hline
		512		& 21	&33.0\\
		1024	& 38	&57.9\\
		2048	& 69	&103.5\\
		4096	& 127	&187.9\\
		8192	&234	&344.6\\
		16384	&434	&637.4\\
		\hline
	\end{tabular}
\end{center}

\subsection{Keys Allocation}

\medskip
\begin{tabular}{lccl}
	\texttt{Public Key}&:&$\bullet$&The $k \times n$ sized matrix $\hat{G}$.\\
	\texttt{Private Keys}&:& $\bullet$ & The matrices $S$ and $P$ of sizes {\small$k\times k$  and $n \times n$} resp.;\\ 
	&& $\bullet$&  The $t-$degree Goppa polynomial $g(z)$ over $\F_{2^m}$, and\\
	&& $\bullet$& The set $L=\{\alpha_1, \alpha_2,\dots, \alpha_n\}\subseteq \F_{2^m}$.
\end{tabular}

The parameters provided in original construction were:\vspace{-2mm}
\begin{center}
	\begin{tabular}{lll}
		$n$ &:&$1024 = 2^{10}$\\
		$t$ &:&$50$\\
		$m$ &:&$10$\\
		$k$ &:&$n-mt = 524$\\
	\end{tabular}
\end{center}
As per given arguments, 
the size of public key is: $$kn = 524 \times 1024 \linebreak = 536576 \text{ bits}$$  $\approx 66$ KB; and \\
The size of private key is: $$(k^2  + n^2) + (t \times m)+ (n \times m)=  (274576 +  1048576)  + 500 + 10240 = 1333892 \text{ bits}$$ $\approx 162.8$ KB.

As the key size in this scheme is very large, Niederreiter proposed the dual variant of McEliece cryptosystem. In that scheme, he made use of dual of the Generalized Reed-Solomon (GRS) codes, $i.e.,$ the parity-check matrix in order to decrease the key size, maintaining same structure.
\bigskip

\section{Niederreiter Cryptosystem}

We describe a variant of the McEliece Cryptosystem published by Harald Niederreiter in  1986 \cite{nie}. Originally this system used GRS codes, which later turned out to be not a suitable choice as the presence of structural attack on them as shown by Sidelnikov and Shestakov \cite{VMSidelnikov} in 1992; then came another Niederreiter's system with the same Goppa codes as used by McEliece. It had the same security as McEliece's system, as shown in \cite{Yuan}. 
Niederreiter's system differs from McEliece's system in public-key structure, encryption mechanism, and decryption mechanism.
In this section, we

The sizes $n, k$ and $t$ are public system parameters, but $g, P$ and $ S$ are randomly generated secrets.

\begin{enumerate}[Step 1:]
	
	\item Firstly \textit{Alice} generates a public and private key pair depending upon publicly available values. During this,
	
	\begin{enumerate}[$(i)$]
		\item Alice selects at random a `$n \times n$' sized permutation matrix $P$;
		
		\item A non-singular `$(n-k) \times (n-k)$' sized matrix $S$;
		
		\item A parity-check matrix $H$ of size `$(n-k) \times n$' for a Goppa code $\Gamma(L,g)$ of dimension $k = n-mt$, where $L =\{\alpha_1,\alpha_2,\dots,\alpha_n\}$ and $g$ is a Goppa polynomial of degree $t$ over $\F_{q^m}$;
		
		\item Publishes her \textbf{Public key} : The $(n-k)\times n$ matrix $S\cdot H \cdot P$;
		
		\item Keeps her \textbf{Private key} : The matrices $P, S$ and $H$.
	\end{enumerate}
	
	\item Suppose $Bob$ has to send a message to $Alice$:
	
	\begin{enumerate}[$(i)$]
		\item Bob has a message \textbf{m} of length `$n$' and Hamming weight `$t$';
		
		\item Computes and send the ciphertext \textbf{c} $ = S\cdot H\cdot P \cdot \textbf{m}^T$.
	\end{enumerate}
	
	\item Suppose $Alice$ receives the ciphertext \textbf{c}:
	
	\begin{enumerate}[$(i)$]
		\item By linear algebra, she finds \textbf{z} such that $H\textbf{z}^T = S^{-1}\textbf{c}$;
		
		\item Applies Patterson's algorithm for decoding of Goppa codes on the vector \textbf{z} to get the codeword $\textbf{z} - \textbf{m}\cdot P^T$, error vector $\textbf{m}\cdot P^T$ and thereby \textbf{m}.
	\end{enumerate}
	
\end{enumerate}

\begin{thm}[\cite{EBerlekamp}]
	If $H \in \F_2^{k \times n}$ and $H \textbf{e} \in \F_2^{n-k}$ (with $wt(\textbf{e})\leq t$) are known, then finding the vector $\textbf{e} \in \F_2^n$ with `$wt(\textbf{e}) \leq t$' is NP complete. Equivalently, general syndrome decoding problem is NP-complete. (Hard problems in Coding Theory)
\end{thm}

\subsection{Equivalence with McEliece PKC}
As mentioned earlier in the text that original Niederreiter scheme worked for Generalized Reed Solomon codes. This lead to a major security threat as mentioned by Sidelnikov and Shestakov \cite{VMSidelnikov}. Though it was earlier stated that the scheme of Niederreiter and McEliece worked alike provided they use same underlying code i.e., binary Goppa code. The benefit of using Niederreiter scheme was that the size of key needed was much reduced. The question of showing both these schemes can be converted into each other, or their equivalence relation is described as below.

\subsection{McEliece to Niederreiter}
In the McEliece PKC, we have a message \textbf{m} of length $k$ and a public key matrix $G'$ of size $k \times n$. We also have an error vector \textbf{e} of length $n$ and weight $t$ which depends on code used. The ciphertext \textbf{c} is obtained as 
$$ \textbf{c}= \textbf{m}G' + \textbf{e}.$$
From our context of coding theory, the parity check matrix can be obtained from generator matrix. Assuming that $G'$ is generator matrix of some code, we obtain $H'$ the parity check matrix for that code.
Now post multiplying $(H')^{T}$ to above relation, we have 
$$ \textbf{c} (H')^{T}= \textbf{m}G'(H')^{T} + \textbf{e}(H')^{T}.$$
Since we have $G' (H')^{T} =0$, we arrive at the following equation
$$ \textbf{c} (H')^{T}= \textbf{e}(H')^{T}.$$
Now the left side of the equation is known as \textbf{c} and $(H')^{T}$ are publicly available. It is also known that weight of \textbf{e} is $t$. Therefore from Niederreiter scheme, we can find the vector \textbf{e}. Once the error vector \textbf{e} is known, going back to original system we have 
$$ \textbf{c}- \textbf{e}= \textbf{m} G'.$$
On expanding the definition of $G'$ we make use of private keys and decoding of Goppa code to get the message \textbf{m}. 
This shows that if Niederreiter scheme is vulnerable/broken then McEliece scheme also suffers.

\subsection{Niederreiter to McEliece}
As per description of Niederreiter scheme, we have message \textbf{y} of length $n$ and weight $t$. The ciphertext \textbf{z} is obtained by multiplication of transpose of an $(n-k) \times n$ sized matrix $H'$ with message, as described
$$ \textbf{z} =  \textbf{y} (H')^T.$$
Using augmented matrix and some facts from linear algebra, one can easily find a vector \textbf{c} of length $n$ having weight at least $t$ such that 
\begin{eqnarray}
\textbf{z}  &=&  \textbf{c} (H')^T \text{ and}\\
\textbf{c} &=&  \textbf{m} G' + \textbf{y}.
\end{eqnarray}
Hence, Niederreiter scheme can be easily converted to McEliece scheme. Thereafter both these schemes hold equivalence in terms of security provided they use same Goppa code.

\subsection{Information-Set Decoding Attack}

Implementation of Information set decoding attack on Niederreiter's scheme can be done in different ways. Like first convert into McEliece's problem and apply same information set decoding algorithm as described in McEliece's scheme. Another method, which is a direct algorithm is described below.

\begin{enumerate}[$(i)$]
	\item Guess an `$n-k$' sized set which  contains all non-zero coordinates of message $\textbf{m}$. The probability of success comes out to be
	$$ \frac{{t \choose t} {n-t\choose n-k-t}}{{n \choose n-k}}.$$
	
	\item Check if the submatrix obtained by the corresponding columns in $S \cdot H \cdot P$ is invertible. The work factor for checking this comes out to be $(n-k)^3$;
	
	\item Pre-multiplying the inverse of this submatrix describes the message vector completely.
\end{enumerate}

In order to execute the attack on this scheme, the work factor comes out to be 
\begin{equation}
\frac{{n \choose n-k}}{{t \choose t} {n-t\choose n-k-t}} \cdot (n-k)^3
\end{equation}
which is the similar as described in McEliece cryptosystem. Hence the corresponding difficulty in applying information-set decoding attack on Niederreiter is same as for McEliece cryptosystem.
\medskip

Niederreiter’s inversion problem is equivalent to McEliece’s inversion problem for the same code. 
In particular, any attack recovering a random e from Niederreiter's $H\textbf{e}$ and $H$ can be used with negligible overhead to recover a random $(\textbf{m},\textbf{e})$ from McEliece's $G\textbf{m} + \textbf{e}$ and $G$. 
Specifically, compute $H$ from $G$, multiply $H$ by $G\textbf{m} + \textbf{e}$ to obtain $HG\textbf{m} + H\textbf{e} = H\textbf{e}$, apply the attack to recover \textbf{e} from $H\textbf{e}$, subtract \textbf{e} from $G\textbf{m} + \textbf{e}$ to obtain $G\textbf{m}$, and recover \textbf{m} by linear algebra.

\subsection{Keys Allocation}

\begin{tabular}{lccl}
	\texttt{Public Key}&:&$\bullet$& The $(n-k)\times n$ sized matrix $S\cdot H \cdot P$.\\
	\texttt{Private Keys} &:& $\bullet$ &Matrices $S$ and $P$ of sizes {\small$(n-k)\times (n-k)$  and $n \times n$} resp.;\\
	&& $\bullet$ & the $t-$degree Goppa polynomial $g(z)$ over $\F_{2^m}$, and\\
	&& $\bullet$& the set $L=\{\alpha_1, \alpha_2,\dots, \alpha_n\}\subseteq \F_{2^m}$.
\end{tabular}

As per the parameters of Goppa codes provided in original construction of McEliece cryptosystem,\vspace{-2mm}
\begin{center}
	\begin{tabular}{lll}
		$n$ &:&$1024 = 2^{10}$\\
		$t$ &:&$50$\\
		$m$ &:&$10$\\
		$k$ &:&$n-mt = 524$,\\
	\end{tabular}
\end{center}
the size of public key is: $$(n-k)n = 500 \times 1024 = 512000 \text{ bits} $$  $\approx 62$ KB; and \\
The size of private key is: $$((n-k)^2  + n^2) + (t \times m)+ (n \times m)=  (250000 +  1048576)  + 500 + 10240 = 1309316 \text{ bits}$$ $ \approx 159.8$ KB.
This shows that the Niederreiter's scheme using dual of Goppa codes can actually decrease the problem of large key size to some extent. 
\bigskip

\section{Classic McEliece: conservative code based cryptography}

A Key Encapsulation Mechanism (KEM) is basically a Public-Key Encryption (PKE) scheme. It consists of three algorithms: Key generation, Encapsulation and Decapsulation. The key generation part is a probabilistic algorithm that takes input a security parameter (like security required in bits) and outputs a public key \texttt{pk} and a private key \texttt{sk}. Secondly, the encapsulation algorithm receives a public key \texttt{pk} and returns a symmetric key (session key) and ciphertext pair $(K, \psi_0)$. Notation wise, $\textnormal{Enc}_{\texttt{pk}}() = (K, \psi_0)$ and $\textnormal{Dec}_{\texttt{sk}}(\psi_0)=K$. Finally, the decapsulation algorithm that receives a private ket \texttt{sk} and a ciphertext $\psi_0$ and outputs either the symmetric key $K$ or failure.

Classic McEliece is a Key Encapsulation Mechanism, which establishes a symmetric key for two end users. 
This KEM is also a candidate in second round \footnote{as on June 2019} for NIST's competition of global standardization of Post Quantum Cryptosystem by Daniel J. Bernstein et al., submitted in $2017$ \cite{CM}. 
It is designed to provide IND-CCA2 security at a very high security level, even against quantum computers. The definition for a KEM to be IND-CCA2 secure follows afterwards.
The KEM is built conservatively from a PKE designed for OW-CPA one way security, namely Niederreiter's dual version of McEliece's public key encryption PKE using binary Goppa codes.
The steps describing KEM are as follows:

\begin{enumerate}[Step 1:]
	\item Suppose $Alice$ asks $Bob$ to establish a session key using Classic McEliece key encapsulation mechanism where the extension field $\F_{2^m}$ is publicly known.

	\item $Bob$ generates his \textit{Classic McEliece key pair} as:
	
	\begin{enumerate}[$(i)$]
		
		\item Firstly, he generates a random monic irreducible polynomial $g(z) \in \F_{2^m}[z]$ of degree `$t$';
		
		\item Selects uniformly random set $\{\alpha_1, \alpha_2, \dots, \alpha_n\} \subseteq \F_{2^m}$ with all distinct elements;
		
		\item Computes a `$t \times n$' sized matrix $\tilde{H} = \left\{h_{i,j}\right\}$ over $\F_{2^m}$, where $h_{i,j}= \alpha_j^{i-1}g(\alpha_i)^{-1}$ for $i=1,2,\dots,t$ and $j=1,2,\dots,n$;
		
		\item Replaces each entry of the matrix $\tilde{H}$ (elements of $\F_{2^m}$) with vectors of $\F_2^m$ (arranged in columns) using vector space isomorphism between $\F_{2^m} $ and $\F_2^m$ to get `$mt \times n$' sized matrix $\hat{H}$;
		
		\item Apply Gaussian elimination on $\hat{H}$ to get a systematic matrix  $H=\left( I_{mt} \mid T_{mt \times (n-mt)}\right)$ if possible, else go back to step $(i)$;
		
		\item Generates a uniform random $n$-bit string $\textbf{s}$;
		
		\item \texttt{\textbf{Public key}:} $T$; and
		
		\item \texttt{\textbf{Private key}:} $\{\textbf{s},g(z),\alpha_1, \alpha_2, \dots, \alpha_n\}$.
		
	\end{enumerate}
	
	\item Using the Public key $T$ of \textit{Bob}, \textit{Alice} starts the \textbf{Key Encapsulation} Process. She generates a session key $K$ and ciphertext $C$ as follows:
	
	\begin{enumerate}[$(i)$]
		\item Generates a uniform random vector $\textbf{e} \in \F_2^n$ with Hamming weight `$t$';
		
		\item \textbf{Niederreiter Encoding:} Using public key `$T$' of \textit{Bob}, first she computes matrix $H$, secondly a vector $C_0 = H \textbf{e}= (I \mid T) \textbf{e}$ of length $mt$ in $ \F_2^{mt}$; 
		
		\item Computes  $C_1$ $= \texttt{H}(2,\textbf{e})$  and generate ciphertext $C=(C_0,C_1)$ of length $mt + 256$ bits;
		
		\item Computes a $256$-bit session key $K= \texttt{H}(1,\textbf{e},C)$. 
		\medskip
		
		Here $\texttt{H}$ is SHAKE256, and the initials $0$, $1$ and $2$ in above hash inputs are represented as a byte.	
	\end{enumerate}
	
	\item \textit{Bob} receives the ciphertext $C$ from \textit{Alice}, and starts decrypting it using \textbf{Decapsulation Process} to generate the some session key $K'$ as:
	
	\begin{enumerate}[$(i)$]
		\item Firstly, he splits $C$ as $(C_0,C_1)$, with $C_0 \in \F_2^{mt}$ and $C_1 \in \F_2^{256}$;
		
		\item Sets $b \leftarrow 1$.
		
		\item \textbf{Decoding step:} 				
		\begin{enumerate}[(a)]
			\item \texttt{Input:} $C_0$ and the private key $\{\textbf{s},g(z),\alpha_1, \alpha_2, \dots, \alpha_n\}$.
			
			\item Extend $C_0$ to $ \textbf{v} = (C_0,0,\dots, 0) \in \F_2^n$ by appending $n -mt$ zeros.
			
			\item Using Niederreiter decoding, find the unique codeword $\textbf{c}$ in Goppa code defined by $$\Gamma =\{~g(z),\alpha_1, \alpha_2, \dots, \alpha_n~\}$$ s.t., $d(\textbf{c},\textbf{v}) \leq t$, if possible. If no such codeword exist, return $\perp$ and move to step $(iii)$.
			
			\item In case when codeword \textbf{c} exists, set the vector $\textbf{e} = \textbf{v} + \textbf{c}$. If wt$(\textbf{e}) = t$ and $C_0 = H\textbf{e}$, return $\textbf{e}$. Otherwise return $\perp.$
		\end{enumerate}
		
		\item If decoding returns $\perp$, set $\textbf{e} \leftarrow \textbf{s}$ and $ b \leftarrow 0$. 
		
		\item Computes $C_1 ' = \texttt{H}(2,\textbf{e})$, and checks if $C_1' = C_1$. If it doesn't match, set $\textbf{e} \leftarrow \textbf{s}$ and $ b \leftarrow 0$.
		
		\item Computes the session key $K' = \texttt{H}(b,\textbf{e},C)$.
		
	\end{enumerate}
	
\end{enumerate}

\textbf{Note: } If there is no failure at any stage during the decapsulation process and $C_1'=C_1$, then surely the session key $K'$ will be identical to $K$.
Equivalently, if Bob receives a valid (legitimate) ciphertext $C$ i.e., $C = (C_0,C_1)$ with $C_0 = H \textbf{e}$ for some $\textbf{e} \in \F_2^n$ of weight $t$ and $C_1 = \texttt{H}(2,\textbf{e})$, the decoding always result in finding the vector \textbf{e}. In this scenario the same session key is established.

\medskip

The generic model of Classic McEliece Key Encapsulation Mechanism is described below:

\medskip
\vspace{-4mm}
\begin{center}
	\boxed{\includegraphics[width=13.75cm, height=9.9cm]{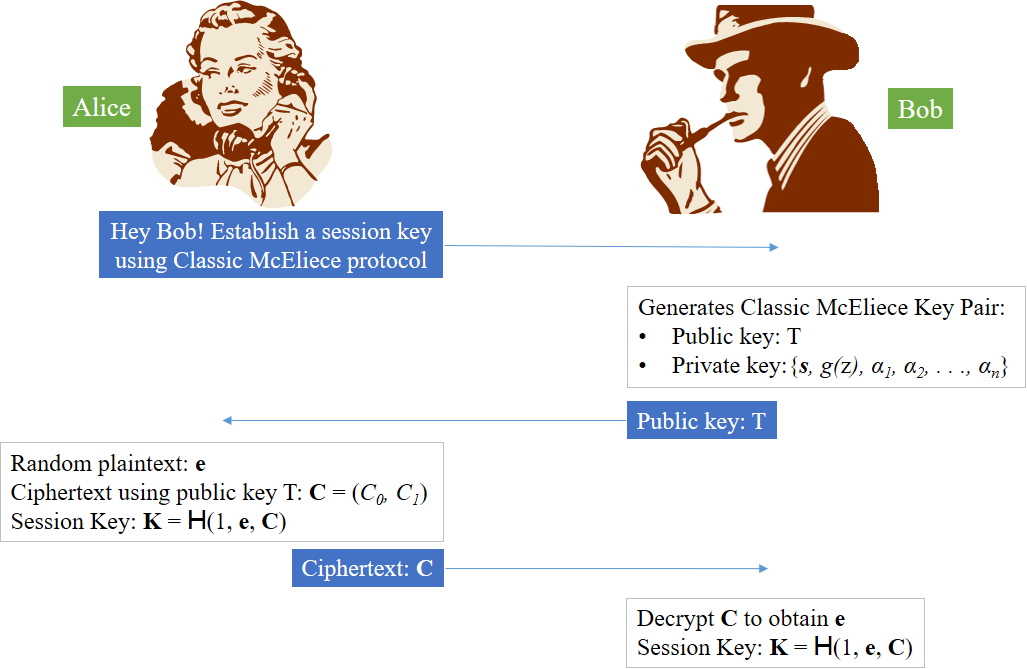}}
\end{center}

\medskip
This representation is the ideal case when Bob has received the ciphertext without any noise and he was able to decrypt that without failure. Otherwise, the same session key is not established.

\begin{itemize}
	\item There is a possibility that Bob computes wrong session key for him. This is possible only if he receives noise in ciphertext received. Further, this can be ensured by both Alice and Bob if 
	
	\begin{itemize}
		\item they communicate and compare hash values of their computed session keys, or
		
		\item Alice transmits $AES_{K}\left(\texttt{H}\left(T\right)\right)$ to Bob so that he can verify if he has got correct session by decrypting the received ciphertext using his session key and matching the result with hash of his public key. 
	\end{itemize}
	If same key is not established, the process re-initiates.
	
	\item It is to be noted that the field $\F_{2^m}$ described in the KEM is defined by a $13$-degree primitive polynomial $$f(z)= z^{13} + z^4 + z^3 + z + 1$$ over $\F_2$ and the Hash function used is SHAKE256 with $32$-byte output. 
	
	\item The values of $n$ and $t$ varies as per different variants.
\end{itemize}
\medskip

From this algorithm, Alice and Bob establish a session key $K$ which can then be used to implement symmetric key cryptography.
Viewing Classic McEliece KEM from coding theory point, we have basically two types of attacks: Decoding attack and Structural attack.
The information-set decoding is the most effective attack strategy known. It does not exploit the structure of generator matrix; it recovers the error vector \textbf{e} from publicly available matrix and the ciphertext.
There are many papers which states algorithms to recover private key from the public key like Sendrier's \textit{support splitting} algorithm. However, despite this and other concerned algorithms, key recovery attacks are vastly slower than information-set decoding.

\begin{example}[Classic McEliece with Small Parameters]
	\textnormal{Suppose Alice asks Bob to establish a Classic \linebreak McEliece based session key. The publicly known extension field $$\F_{2^4} = \frac{\F_2[x]}{\left<x^4+x^3+1 \right>} = \F_2(\beta),$$ where $\beta$, being a primitive element of the field $\F_{2^4}$, is a root of the irreducible polynomial $x^4+x^3+1$.
		Then the field $\F_{2^4} = \F_2(\beta)$ consists of $\{0,1,\beta, \beta^2,\dots, \beta^{14}\}$. Here $\beta^{15} = 1$.
		The error correcting capacity i.e., the parameter $t =2$ is also publicly known.
		\begin{center}
			\textbf{Key Generation}
		\end{center}	
		\begin{enumerate}[$(i) $]
			\item Bob generates $g(z)=z^2+z+\beta$ an irreducible polynomial over  $\F_2(\beta)$ of degree $t= 2 $.
			\item Let $L= \F_2(\beta) = \{0,1,\beta, \beta^2,\dots, \beta^{14}\}$ be set of $n=16$ elements of $\F_2(\beta)$.\\
			We make use of the following table to describe elements of the field $\F_2(\beta)$:
			\begin{center}
				$ \begin{array}{ccllllc}
				0&=&&&&&=(0,0,0,0)^T;\\
				1& =&  1 & & & & = (1,0,0,0)^T;\\
				\beta & = & &\beta & & & = (0,1,0,0)^T;\\
				\beta^2 & = & & &\beta^2 & & = (0,0,1,0)^T;\\
				\beta^3 & = & & & &\beta^3 & = (0,0,0,1)^T;\\
				\beta^4 & = &1+ & & & \beta^3& = (1,0,0,1)^T;\\
				\beta^5 & = &1+ &\beta+ & &\beta^3 & = (1,1,0,1)^T;\\
				\beta^6 & = &1+ &\beta+ &\beta^2+ &\beta^3 & = (1,1,1,1)^T;\\
				\beta^7 & = &1+ &\beta+ &\beta^2 & & = (1,1,1,0)^T;\\
				\beta^8 & = &   &\beta+ &\beta^2+ &\beta^3 & = (0,1,1,1)^T;\\
				\beta^9 & = &1+ & &\beta^2 & & = (1,0,1,0)^T;\\
				\beta^{10} & = & &\beta+ & &\beta^3 & = (0,1,0,1)^T;\\
				\beta^{11} & = &1+ & &\beta^2+ & \beta^3& = (1,0,1,1)^T;\\
				\beta^{12} & = &1+ & \beta& & & = (1,1,0,0)^T;\\
				\beta^{13} & = & &\beta+ & \beta^2&& = (0,1,1,0)^T;\\
				\beta^{14} & = &  & &\beta^2+ & \beta^3& = (0,0,1,1)^T;\\
				\end{array}$
			\end{center}
			\item As described in the algorithm, 
			\begin{itemize}
				\item $h_{1,1} = (1+1+\beta)^{-1} = \beta^{-1} = \beta^{14}$;
				\item $h_{1,2} = (0+0+\beta)^{-1} = \beta^{-1} = \beta^{14}$;
				\item $h_{1,3} = (\beta^2 + \beta + \beta)^{-1}= \beta^{-2} = \beta^{13}$;
				\item $h_{1,4} = (\beta^4 + \beta^2 + \beta)^{-1} = (\beta^3+ \beta^2 + \beta + 1)^{-1} = (\beta^6)^{-1} = \beta^9$;
				\item $h_{1,5} = (\beta^6 + \beta^3 + \beta)^{-1} = \beta^6$, etc.
				\item $h_{2,1} = 0 \cdot (1+1+\beta)^{-1} = 0$;
				\item $h_{2,2} = 1 \cdot (0+0+\beta)^{-1} = \beta^{-1} = \beta^{14}$;
				\item $h_{2,3} = \beta \cdot (\beta^2 + \beta + \beta)^{-1}= \beta \cdot \beta^{-2} = \beta^{-1} = \beta^{14}$;
				\item $h_{2,4} = \beta^2 \cdot (\beta^4 + \beta^2 + \beta)^{-1} = \beta^2 \cdot \beta^9 = \beta^{11}$;
				\item $h_{2,5} = \beta^3 \cdot \beta^6 = \beta^9$, etc.
			\end{itemize}
			The $2 \times 16$ sized matrix $\tilde{H}$ is given by 
			{\small\begin{equation*}\left( \begin{array}{cccccccccccccccc}
				\beta^{14}&\beta^{14}&\beta^{13}&\beta^9&\beta^{6}&\beta^6&\beta^{3}&\beta^{7}&\beta^{11}&\beta^{7}&\beta^{9}&\beta^{3}&\beta^{12}&\beta^{13}&\beta^{11}&\beta^{12}\\
				0& \beta^{14}& \beta^{14}& \beta^{11}& \beta^9& \beta^{10}& \beta^8& \beta^{13}& \beta^3& 1& \beta^3& \beta^{13}& \beta^8& \beta^{10}& \beta^9& \beta^{11}\\
				\end{array}
				\right)\end{equation*}}
			\item The matrix $\hat{H}$ becomes
			\begin{equation*}\hat{H} =\left( \begin{array}{cccccccccccccccc}
			0&0&0&1&1&1&0&1&1&1&1&0&1&0&1&1\\
			0&0&1&0&1&1&0&1&0&1&0&0&1&1&0&1\\
			1&1&1&1&1&1&0&1&1&1&1&0&0&1&1&0\\
			1&1&0&0&1&1&1&0&1&0&0&1&0&0&1&0\\
			0&0&0&1&1&0&0&0&0&1&0&0&0&0&1&1\\
			0&0&0&0&0&1&1&1&0&0&0&1&1&1&0&0\\
			0&1&1&1&1&0&1&1&0&0&0&1&1&0&1&1\\
			0&1&1&1&0&1&1&0&1&0&1&0&1&1&0&1\\
			\end{array} \right).
			\end{equation*}
			\item On applying Gaussian elimination, the matrix $H$ becomes
			\begin{equation*}H =\left( \begin{array}{cccccccc|cccccccc}
			1&0&0&0&0&0&0&0&1&0&0&0&0&1&1&1\\
			0&1&0&0&0&0&0&0&0&1&1&1&0&0&0&1\\
			0&0&1&0&0&0&0&0&1&1&1&1&1&0&1&1\\
			0&0&0&1&0&0&0&0&0&1&0&1&1&1&0&1\\
			0&0&0&0&1&0&0&0&0&0&0&1&1&1&1&0\\
			0&0&0&0&0&1&0&0&1&1&0&0&1&1&1&0\\
			0&0&0&0&0&0&1&0&1&0&1&1&0&1&0&0\\
			0&0&0&0&0&0&0&1&0&1&1&0&0&1&1&0\\
			\end{array} \right).
			\end{equation*}
			\item Let the random 16-bit string $\textbf{s}= (0000000000000000)$.
			\item The Public key is given by
			\begin{equation*}T =\left( \begin{array}{cccccccc}
			1&0&0&0&0&1&1&1\\
			0&1&1&1&0&0&0&1\\
			1&1&1&1&1&0&1&1\\
			0&1&0&1&1&1&0&1\\
			0&0&0&1&1&1&1&0\\
			1&1&0&0&1&1&1&0\\
			1&0&1&1&0&1&0&0\\
			0&1&1&0&0&1&1&0\\
			\end{array} \right).
			\end{equation*}
			The private key is $(\textbf{s}, g(z), \F_2(\beta))$.
		\end{enumerate}
		\begin{center}
			Now, Bob shares his public key $T$ with Alice. On receiving this matrix Alice does \\Key Encapsulation process as described
			\textbf{Key Encapsulation}
		\end{center}
		\begin{enumerate}[$(i)$ ]
			\item Generate random plaintext $\textbf{e}=(1100000000000000)$ of length $n=16$ and weight $t=2$.
			\item $C_0 =$ $H \textbf{e}= (11000000)$.
			\item $C_1=\texttt{H}(2,\textbf{e})= 26fe36f811ac8fe9f19ba997a39d3682ef06b29509cca1903ffe4a0b247c833f =$\\
			00100110111111100011011011111000000100011010110010001111111010011111000110011011101$\linebreak$
			01001101011110100011100111010011011010000010111011110011010110010100101010100111001$\linebreak$ 10010100001100100011111111111110010010100101100100100011111001000001100111111.$\linebreak$ Assuming hash to be SHA256.
			\item Session key $K= \texttt{H}(1,\textbf{e},C)=$\\
			90d7c9dccc4689f6894b1b6e58ee9b3832 8e4df9937536eb9b5715a38ee4e1be. The output ciphertext \\$C=(C_0,C_1)=$\\ 11000000001001101111111000110110111110000001000110101100100011111110100111110001100\\
			11011101010011001011110100011100111010011011010000010111011110011010110010100101010\\
			10011100110010100001100100011111111111110010010100101100100100011111001000001100111\\111.
		\end{enumerate}
		On receiving this ciphertext, Bob performs decapsulation process. During this, he splits the first 8 bits as \\$C_0$, executes decoding on it to find $\textbf{e}$. This process is explained below with reference to Proposition 1.6.1.:
		\begin{center}
			\textbf{Decapsulation}
		\end{center}
		\begin{enumerate}[$(i)$ ]
			\item The vector $\textbf{v} = (C_0, 00000000) = (1100000000000000)$.
			\item Finding nearest codeword in Goppa code $\Gamma$:
			\begin{itemize}
				\item Syndrome of the received vector \textbf{v} is calculated as $$S(\textbf{v}) = \frac{1}{z} + \frac{1}{z+1} \mod g(z).$$
				Hence $S(\textbf{v})= \beta^{13}$.
				\item The key equation $S(\textbf{z})\sigma(z) \equiv w(z) \mod g(z)$ implies
				$$(z^2 + (\alpha_1 + \alpha_2)z+ \alpha_1 \alpha_2)\beta^{13} = \alpha_1 + \alpha_2 \mod g(z).$$
				\item Thus, on comparing the coefficients both sides, we get $$\alpha_1 + \alpha_2 =0 \text{ and } \alpha_1 \alpha_2 =0.$$
				\item Hence $\alpha_1 =0$ and $\alpha_2 = 1$.
				\item Thus the error in received vector becomes $(1100000000000000)$. This means the closest codeword $\textbf{c} = (0000000000000000)$.
				\item The vector $\textbf{e}= \textbf{v}+\textbf{c} = (1100000000000000)$.
			\end{itemize}
			\item As the Hamming weight of vector \textbf{e} is equal to $t =2$, and $C_1'= \texttt{H}(2,\textbf{e})= C_1$, Bob computes the session key $K=\texttt{H}(1,\textbf{e},C)$.
		\end{enumerate}
	}\end{example}
 $\hfill \Box$

\subsection{Information-Set Decoding Attack}

The Classic McEliece key encapsulation mechanism submitted in NIST contains two variants depending upon parameter sets which provides different security levels.

\begin{center}
	\begin{tabular}{|l|c|c|c|c|c|}
		\hline
		\textcolor{blue}{Variant} 		& \textcolor{blue}{$n$}  & \textcolor{blue}{$m$} & \textcolor{blue}{$t$} & \textcolor{blue}{$k=n-mt$} & \textcolor{blue}{Security}\\
		\hline
		\textcolor{red}{mceliece6960119} & 6960 & 13  & 119 & 5413	  & 128 \\ 
		\hline
		\textcolor{red}{mceliece8192128} & 8192 & 13  & 128 & 6528     & 256\\
		\hline
	\end{tabular}
\end{center}	
\vspace{2mm}
Classic McEliece works by operating dual of the McEliece cryptosystem (Niederreiter cryptosystem). Both the systems provide same security, and the information-set decoding also works in same fashion. Suppose the attacker has received the ciphertext $(C_0,C_1)$, then he applies information set-decoding on $C_0$ as he knows the length of $C_0$ and since $C_0= H \textbf{e}$.

As per the work factor described in Niederreiter's information-set decoding attack,
\begin{equation}
(n-k)^3 \cdot \frac{{n \choose n-k}}{{n-t \choose n-k-t}}.
\end{equation}
For the first variant \textbf{mceliece6960119} parameters used are $n=6960$, $m=13$, $t=119$ and $k=n-mt=5413$. The work factor comes out to be 
$$ 7.5 \times 10^{88} \approx 2^{295}.$$

For the second parameters set viz. \textbf{mceliece8192128}, the work factor for implementation of information-set decoding attack comes out to be
$$ 1.03 \times 10^{100} \approx 2^{332}.$$

\textbf{Using McEliece conversion:}\\
\medskip

After converting Niederreiter scheme into McEliece scheme, the information set decoding parameters are changed as the work factor becomes
\begin{equation}
k^3 \cdot \frac{{n \choose k}}{{n-t \choose k}}.
\end{equation}
Now plugging-in, the parameters set $n=6960$, $m=13$, $t=119$ and $k=n-mt=5413$, the work factor comes out to be $$\approx 3.2 \times 10^{90} \approx 2^{301}.$$
Further, for the parameters set $n=8192$, $m=13$, $t=128$, the work factor becomes
$$\approx 6 \times 10^{101} \approx 2^{338}.$$

In the paper of Daniel J. Bernstein et al. \cite{DJBLength} in 2008, it is proved that the number of bit operations to break the $(6960,13,119)$ variant is $2^{266.94}$.

\subsection{Chosen-Ciphertext Attacks}

For this Key Encapsulation Mechanism, chosen ciphertext attacks do not work as per following reasons.
\begin{enumerate}[$\star$]
	\item Ciphertext includes hash of the message as a confirmation, and the attacker can never compute the hash of a modified version of message without knowing message in the first place.
	\item There are no decryption failures, i.e., the modified ciphertext will produce an unpredictable session key, whether or not the modified message vector has weight $t$.
\end{enumerate}

\subsection{Keys Allocation}

\begin{tabular}{lccl}
	\texttt{Public Key}&:& $\bullet$&The $mt \times (n-mt)$ sized matrix $T$.\\
	\texttt{Private Keys}&: & $\bullet$ & $\textbf{s} \in \F_2^n$; \\
	&&$\bullet$& the $t$-degree Goppa polynomial $g(z)$ over $\F_{2^m}$, and\\
	&&$\bullet$& the set $L=\{\alpha_1, \alpha_2,\dots, \alpha_n\}\subseteq \F_{2^m}$.
\end{tabular}

Here we examine the key sizes as per the different variants proposed. These are expressed as:

\subsection{mceliece6960119}

Each row of $T$ is represented as $\lceil (n-mt)/8 \rceil$-byte string, and there are total $mt$ rows in $T$. As per given parameters, the size of public key is:
$$mt \lceil(n-mt)/8\rceil  =  13 \cdot 119 \cdot \lceil(6960-13 \cdot 119)/8\rceil = 1047319 \text{ bytes} $$ $\approx 1$ MB.\\

Private key consist of $(\textbf{s}, g(z), \alpha_1, \dots, \alpha_n)$. For representation of string $\textbf{s} \in \F_2^n$, $\lceil n/8 \rceil$-bytes are stored. The polynomial $g(z)$ is a monic polynomial of degree $t$ over $\F_{2^m}$. Hence it saves $t$ coefficients where each coefficient is $\lceil m/8 \rceil$-byte string. So total there are $t \lceil m/8 \rceil$-bytes for saving this. The sequence of elements $(\alpha_1, \alpha_2,\dots,\alpha_n)$ are stored as $n$ field elements using Bene$\hat{\textnormal{s}}$ network.
The size of private keys is: \\
$\lceil n/8 \rceil + t \lceil m/8 \rceil + \lceil (2m-1)2^{m-4} \rceil   = 870+ 238+ 12800 = 13908 \text{ bytes}$ $\approx 13.6$ KB. 

\vspace{-4mm}\subsection{mceliece8192128}
Similarly, for this set of parameters, the size of public key is: $$mt \lceil(n-mt)/8\rceil mt = \lceil(8192-13 \cdot 128)/8\rceil \cdot 13 \cdot 128 = 1357824 \text{ bytes}\approx 1.3 \text{MB}.$$
The size of private keys: $\lceil n/8 \rceil + t \lceil m/8 \rceil + \lceil (2m-1)2^{m-4} \rceil$ (using Bene$\hat{\textnormal{s}}$ network)  $= 1024+ 256+ 12800 = 14080 \text{ bytes}$ $\approx 13.75$ KB.

\bigskip


\section{Strength of the Cryptosystem}

Clearly brute force is not feasible to all the discussed cryptosystems.	However, original McEliece cryptosystem is vulnerable to chosen-plaintext attacks. 
The encoding matrix is the public key, usually publicly available, and the attacker can simply guess some plaintext, construct the corresponding ciphertext and compare this to the target ciphertext. 
This system, based up on original parameters, is now completely broken.
However, on suitably increasing the size of parameters, the scheme is proved to resist all kind of attacks. The standard classification of cryptographic attacks to block ciphers concording the amount and quality of secret information they are able to discover are listed as:
\begin{itemize}
	\item \textbf{Total break} - the attacker deduces the secret key.
	\item \textbf{Global deduction} - the attacker discovers an equivalent algorithm for encryption and decryption without learning the secret key.
	\item \textbf{Local deduction} - the attacker discovers additional plaintext-ciphertext which were not earlier known.
	\item \textbf{Information deduction} - the attacker gains some Shannon information about plaintext-ciphertext pairs.
	\item \textbf{Distinguishing algorithm} - the attacker can distinguish the cipher from a random string.
\end{itemize}
The McEliece PKC is immune to total break in polynomial time. 
However, the original system is vulnerable to chosen-ciphertext attack. 
Suppose message \textbf{m} is encrypted as $\textbf{c}= \textbf{m} G + \textbf{e}$. 
If we select one bit at random from the set of bits which corresponds to 1, and one bit from the set of bits which correspond to 0 in the error vector \textbf{e}, then upon inverting these bits, we get a different ciphertext having $t$ errors. 
With probability ${t(n-t)}/{n(n-1)}$, a new different ciphertext will be produced containing exactly $t$ errors.
Thus assuming a decryption oracle is available, the attacker sends this new ciphertext to the decryption oracle that will output the plaintext, breaking the system. 

As per the original parameters of McEliece PKC i.e., for $n=1024$, and $t=50$ the probability for choosing different ciphertext with corresponding to same message as described above becomes
\begin{equation}
\frac{t(n-t)}{n(n-1)} = \frac{50 \cdot 974}{1024 \cdot 1023} = 0.046489.
\end{equation}
Hence the work factor becomes 21.5. So out of 22 calls to decryption oracle, the attacker can implement adaptive CCA attack on this PKC.

\begin{dfn}
	\textnormal{\cite{EPersichetti} The adaptive Chosen-Ciphertext Attack game for a KEM proceeds as follows:}
	\begin{enumerate}[$(i)$]
		\item Query a key generation oracle to obtain a public key \texttt{pk}.
		
		\item Make a sequence of calls to a decapsulation oracle, submitting any string of the proper length. Oracle will respond the result after decapsulation of this string.
		
		\item Query an encapsulation oracle. The oracle runs the encapsulation algorithm and produces a pair $(\bar{K}, \bar{\phi_0})$ and a random a random string $K^*$ of same length as of $K$. Then oracle replies the challenger both pairs $(\bar{K}, \bar{\phi_0})$ say if $b=0$ and $(\bar{K^*}, \bar{\phi_0})$ if $b=1$.
		
		\item Challenger then keep performing decapsulation queries for strings other than those challenged above.
		
		\item Challenger outputs $b^* \in \{0,1\}$. 
	\end{enumerate}

	The adversary succeeds if $b^*=0$ i.e., when it corresponds to correct pair $(\bar{K}, \bar{\phi_0})$. More precisely, we define the advantage $\mathcal{A}$ against KEM as 
	$$ Adv_{KEM}(\mathcal{A}, \lambda)= \left| Pr[b^*= 0 ] - \frac{1}{2} \right|.$$
	We say that a KEM is secure under adaptive chosen ciphertext attacks if the advantage $Adv_{KEM}$ of any polynomial time adversary $\mathcal{A}$ in the above CCA model is negligible.
\end{dfn}
$\linebreak$McEliece's original PKE was not designed to resist chosen-ciphertext attacks, but the KEM Classic McEliece possesses IND-CCA2 security. 
It employs the best practices manifested like:
\begin{itemize}
	\item The session key comes from hash of uniform random input vector $\textbf{e}$.
	\item Ciphertext consists of confirmation also, ie., another hash of vector $\textbf{e}$.
	\item After computation of $\textbf{e}$, using private key, from ciphertext, ciphertext is recomputed for confirmation that it matches.
	\item If decryption fails for reverse computation, KEM do not return failure; instead it return a pseudo-random function of the ciphertext, specifically a cryptographic hash of a separate private key and the ciphertext.
\end{itemize}

Subsequently there have been a lot of publications studying the one-wayness of the system and introducing sophisticated non-quantum attack algorithms:
Clark-Cain \cite{clark}, crediting Omura; Lee-Brickell \cite{lee}; Leon \cite{leon}; Krouk \cite{krouk}; Stern \cite{stern}; van Tilburg \cite{tilburg}; Chabaud \cite{chabaud}; Bernstein-Lange-Peters \cite{blp}; Finiasz-Sendrier \cite{FS}; May-Meurer-Thomae \cite{mmt}. 

\medskip
This led to transformation of McEliece cryptosystem to public key cryptosystem, namely Classic McEliece.
It is structured in the Niederreiter's dual version of the McEliece scheme. 
It is a key encapsulation mechanism designed to exchange the symmetric key using public key cryptosystem.
Ciphertext includes hash function values to provide integrity check.
The indistinguishability criteria against chosen ciphertext attacks for a KEM is elaborated as follows.

\medskip
In order to be secure against adaptive chosen-ciphertext attacks, for a query of getting plaintext from random ciphertext, there are no decryption failures. 
For a non legitimate (invalid) ciphertext, the decapsulation process work and outputs some session key.
The KEM is structured in such a way that it do not leak side channel errors in case of decryption failures. 

\medskip
With the same key-size optimizations, the Classic McEliece system uses a key size of $(c_0 + o(1))b^2(\log b)^2$ bits to achieve $2^b$ security against all non-quantum attacks known today, where $c_0$ is exactly the same constant. 
All of the improvements have disappeared into the $o(1)$.
The decapsulation process does not reveal any additional information: i.e., all attacks are as difficult as passive attacks. It outputs a session key for all ciphertexts whether they are valid or not.
To be precise in applying information set decoding, a random set of $k$ positions to be an information set with reasonable probability is $29 \%$. However, the chance of this set being error free drops rapidly as the number of errors increase. 

\medskip
Some applications for iPhone and iPad use McEliece public key encryption scheme are the S2S application and PQChat. 
\begin{itemize}
	\item In S2S app, files are encrypted with users' public keys and stored in the cloud so that they may be shared. Sharing is by means of links that index the encrypted files on the cloud and each user uses their private key to decrypt the shared files.
	\item The PQChat is a secure instant messaging system application which uses McEliece cryptosystem to provide security.
\end{itemize}

The first parameters set for Classic McEliece ``mceliece6960119'' takes $m =13$, $n = 6960$ and $t =119$. This parameters set came when the original McEliece parameters $(10,1024,50)$ were proved to be attacked.
The subsequent information set decoding have marginally reduced the number of bit operations considerably below $2^{256}$.

\medskip

Concerning efficiency, the use of random-looking linear codes with no visible structure draws public-key sizes to be on the scale of a MB for quantitatively high security: the public key is a full (generator/parity-check) matrix. 
Applications must extend using each public key for long enough to handle the costs of generating and distributing the key.
\bigskip

\section{Conclusion \& Future Work}

Based on hard problems which do not seem to be affected by presently available quantum algorithms, there is a scope for McEliece cryptosystem based scheme to be used as post-quantum cryptosystem candidate. 
The duration of its analysis also contribute to the security of McEliece PKC since it is as former as RSA. 
The problem of handling and operating with large keys is the only concern with this cryptosystem.
Although, there is always a scope of reducing the key size further without giving a loss to its security.

\medskip
In this report, we focused on a public key cryptosystem, which is designed long back based on coding theory which can be used in practice once large scale quantum computers are built.
It is expected that breaking Classic McEliece with parameters $(6960, 119)$ is more expensive than to break AES$-256$ in both pre-quantum and much more in post-quantum scenario. 
The future work for this cryptosystem would be to implement the attacks on it and to compare this system with other post-quantum cryptosystems. 

\medskip
We have compiled MATLAB code to execute the implementation of information set decoding attack for a given generator matrix for a linear code over binary field. Accompanying that, we have added a method to find generalized inverse of any matrix over binary field, which is somehow useful in context as described in following Appendix.

\medskip

In future, we will try formulating algorithms based on quantum computing for information set decoding or any attack which can be implemented on this system.
Moreover,  when analyzing other code based post quantum cryptosystems we can compare their structure with this system and find possibilities of further improvements or attacks.

\pagebreak

{\large \textbf{Acknowledgment}}

\medskip
The author is thankful to Ms Pratibha Yadav for encouraging him to work in this area and also wish to thank Mr Amit Kumar for helping in formulating the codes for generalized inverses and information set decoding attacks. Finally, the author is grateful to Dr. Dhananjoy Dey for carefully reading the manuscript and making many valuable corrections and suggestions to improve this report.

\bigskip
{\wh $\cdot$}
\bigskip

\appendix
{\Huge \textbf{Appendix}}
\begin{appendices}
	\section{Generalized Inverses}
	The world of mathematics revolve about a few key equations. One of the famous problems of mathematics are solving the following
	\begin{equation}
	\textbf{Ax}=\textbf{b}, \text{ where } \textbf{A} \in \mathbb{C}^{m \times n}, \textbf{x} \in \mathbb{C}^n \text { and } \textbf{b} \in \mathbb{C}^m.
	\end{equation}
	This type of problems appear in many abstract or arithmetic cases. We have a well known result that 
	\begin{itemize}
		\item if $rank[\textbf{A}:\textbf{b}] = rank(\textbf{A})$, then 
		\begin{itemize}
			\item there is a unique solution if $rank(\textbf{A}) = n$; 
			\item there are infinitely many solutions if $rank(\textbf{A}) < n$;
		\end{itemize}
		\item there is no solution if $rank[\textbf{A} : \textbf{b}] \neq rank(\textbf{A})$.
	\end{itemize}
	
	An obvious case if the matrix \textbf{A} is square and its determinant is non-zero, the solution to above system would be $\textbf{x} = \textbf{A}^{-1} \textbf{b}$.
	Here we talk about inverse of a matrix. The inverse exists only when matrix is invertible i.e., non-singular. Now we move towards generalized inverses or Moore-Penrose inverse of any matrix.
	
	\begin{dfn}[Generalized inverse]
		\textnormal{For any matrix $\textbf{A}\in \mathbb{C}^{m \times n}$, the generalized inverse of \textbf{A}, denoted by $\textbf{A}^{\dagger}$ is a unique matrix in $\mathbb{C}^{n \times m}$ such that
			\begin{enumerate}[$(i)$ ]
				\item $\textbf{AA}^{\dagger}\textbf{A} = \textbf{A}$,
				\item $\textbf{A}^{\dagger}\textbf{A}\textbf{A}^{\dagger} = \textbf{A}^{\dagger}$,
				\item $\left(\textbf{A}\textbf{A}^{\dagger}\right)^*=\textbf{A}\textbf{A}^{\dagger}$,
				\item $\left(\textbf{A}^{\dagger}\textbf{A}\right)^*=\textbf{A}^{\dagger}\textbf{A}$.
		\end{enumerate}}
	\end{dfn}
	
	We will see a few interesting properties of generalized inverse once we compute it. We make use of following proposition for achieving that aim.
	
	\begin{proposition}
		If $\textbf{A} \in \mathbb{C}^{m \times n}$, then there exists $\textbf{B} \in \mathbb{C}^{m \times r}$ and $\textbf{C} \in \mathbb{C}^{r \times n}$ such that
		\begin{eqnarray}
		&\textbf{A} = \textbf{BC} \text{ and }\\
		&rank(\textbf{A})=rank(\textbf{B})=rank(\textbf{C})=r.
		\end{eqnarray}
	\end{proposition}
	The proof of this result follows by taking into account the Echelon form of \textbf{A}. The echelon form of matrix \textbf{A} if of the form $$ \textbf{E}_{\textbf{A}}= \left[ 
	\begin{matrix}
	\textbf{C}_{r\times n} \\
	\textbf{0}_{(m-r) \times n}
	\end{matrix}
	\right]$$
	This shows how the matrix \textbf{C} is generated.
	The echelon form of \textbf{A} is obtained by applying permutations to rows of \textbf{A}, this means $\textbf{E}_{\textbf{A}} = \textbf{P A}$. So the matrix \textbf{B} is inverse of the permutation matrix \textbf{P}. For further proof and readings, one may refer \cite{geninv}. Using this method one may compute the decomposition matrices having aforementioned properties. These contribute to computing the generalized inverse as following result.
	
	\begin{thm}\cite{geninv}
		If $\textbf{A}=\textbf{BC}$ where $\textbf{A} \in \mathbb{C}^{m \times n}, \textbf{B} \in \mathbb{C}^{m \times r}, \textbf{C} \in \mathbb{C}^{r \times n}$ and $r = rank(\textbf{A})=rank(\textbf{B})=rank(\textbf{C})$, then \begin{equation}
		\textbf{A}^{\dagger} = \textbf{C}^*(\textbf{CC}^*)^{-1}(\textbf{B}^*\textbf{B})^{-1}\textbf{B}^*.
		\end{equation}$\hfill \Box$
	\end{thm}
	
	This matrix have certain useful properties. Firstly, it may be noted that matrices $\textbf{B}^*\textbf{B}$ and $\textbf{CC}^*$ are of rank $r$ in $\mathbb{C}^{r \times r}$. Hence these are invertible matrices.
	If the given matrix \textbf{A} is already an invertible matrix, the generalized inverse $\textbf{A}^{\dagger}$ is identical to inverse of \textbf{A} i.e., $\textbf{A}^{-1}$.
	
	\begin{cor}
		As a consequence to this, we have 
		\begin{eqnarray}{\label{a5}}
		\textbf{AA}^{\dagger}&=& \textbf{BCC}^*(\textbf{CC}^*)^{-1}(\textbf{B}^*\textbf{B})^{-1}\textbf{B}^* = (\textbf{B}^*\textbf{B})^{-1}\textbf{B}^*\\
		\textbf{A}^{\dagger}\textbf{A}&=& \textbf{C}^*(\textbf{CC}^*)^{-1}(\textbf{B}^*\textbf{B})^{-1}\textbf{B}^*\textbf{BC} =\textbf{C}^*(\textbf{CC}^*)^{-1}.
		\end{eqnarray}
	\end{cor}
	
	Now, anyhow on computing matrices \textbf{B} and \textbf{C}, if it comes out that either of them is an identity matrix, we have a crucial result.
	
	If the matrix \textbf{B} is identity, then in (\ref{a5}), $\textbf{AA}^{\dagger} = \textbf{I}$. Similar result holds for other case.
	
	\subsection{Code-based Cryptography}
	
	The generator matrices of codes are prone to generalized inverses since for an $(n,k)-$code, the generator matrix is a $k \times n$ matrix of rank $k$ over finite field. 
	For such matrices, if Proposition A.0.1 holds true and the matrices $(\textbf{CC}^*)$ and $(\textbf{B}^*\textbf{B})$ are invertible in the underlying field, the generalized inverse exists.
	Wu C.K. et al. in 1998 \cite{wu} showed that for any matrix over finite field, its $\{1,2\}-$inverse always exist. For $\{1,2\}-$inverse, it means that the points $(i)$ and $(ii)$ of Definition A.0.2 are only satisfied by the ``pseudo-generalized inverse''. 
	
	Similarly for parity check matrices, generalized inverse exist.  In case of Classic McEliece KEM, the public key is indeed the parity check matrix for the Goppa code and is of special kind
	$$\textbf{H}=\left(\textbf{I}_{mt} \mid \textbf{T}_{mt \times (n-mt)}\right).$$
	In this scenario, the decomposition matrices from Proposition A.0.1 comes out to be $\textbf{B}= \textbf{I}_{mt}$ and $\textbf{C} = \textbf{H}$. Hence as per Corollary A.0.1.1, $\textbf{H} \textbf{H}^{\dagger} = \textbf{I}$ if the matrix $\textbf{HH}^T$ is invertible in $\F_2$.
	On making use of this notion, if we had a message \textbf{m}, which  was to be encrypted as \textbf{mH}, where \textbf{H} is the publicly known matrix of above mentioned type, the decryption would be simply multiplying by generalized inverse of \textbf{H} i.e., $\textbf{mHH}^{\dagger}$.
	Since the message is encrypted by post multiplication with transpose of \textbf{H}, it is prone towards this attack. The amount of work needed to be done in order to find $\textbf{H}^{\dagger}$ will be same as finding information set for the same matrix. 
	
	\subsection{MATLAB code for Generalized Inverses of Generator Matrix}

	\subsection*{Reading Generator Matrix B from file `Bmatrix.txt' over binary field}
	
	\begin{verbatim}
	fid = fopen(`Bmatrix.txt');
	Bdim = fscanf(fid,`%d',[1 2]);
	B  = fscanf(fid,`%d',[Bdim(1,1),Bdim(1,2)]);
	fclose(fid);
	
	m = Bdim(1,1); % Num. of rows of B
	n  = Bdim(1,2); % Num. of columns of B
	A = B;
	\end{verbatim}

	\subsection*{Echelon Matrix Preparation}
	
	\begin{verbatim}
	if m<n
	mindim = m;
	maxdim = n;
	else
	mindim = n;
	maxdim = m;
	end
	counter = maxdim;
	flag = 0;
	j = 1;
	I = eye(m);
	for i= 1:counter
	if A(j,i) ~= 1
	for k = (j+1):m
	if A(k,i) == 1
	A(j,:) = mod((A(j,:) - A(k,:)),2);
	I(j,:) = mod((I(j,:) - I(k,:)),2);
	flag = 1;
	break;
	end
	end
	else
	flag = 1;
	end
	
	if flag == 0
	if(j >=m || i>=n)
	break;
	end
	continue;
	else
	for k = (j+1):m
	if A(k,i) ~= 0
	A(k,:) = mod((A(k,:) - A(j,:)),2);
	I(k,:) = mod((I(k,:) - I(j,:)),2);
	end
	end
	j = j +1;
	
	if j>m
	break;
	end
	end
	flag = 0;
	end
	
	C = mod(inv(I),2); % Inverse of permutation Matrix
	\end{verbatim}

	\subsection*{Calculation of $\textbf{B}^{\dagger}$:}
	
	\begin{par}
		BDagg = A' * inv(A*A') * inv(C'*C) * C'
	\end{par} \vspace{1em}
	\begin{verbatim}
	if (mod(det(A*A.'),2) ~= 0 && mod(det(C.'*C),2) ~=0)
	BDagg = (A.')*det(A*A.')*(inv(A*A.'))*det(C.'*C)*(inv(C.'*C))*C.';
	BDagg = mod(int32(BDagg),2);
	BDagg = double(BDagg);
	
	fprintf(1,`\nOriginal Matrix :');
	B
	fprintf(1,`\nGeneralized Inverse Matrix :');
	BDagg
	
	%Relations Check
	% Relation 1. B*BDagg*B = B
	if mod(B*BDagg*B,2) == mod(B,2)
	fprintf(1,`\nRelation B*BDagg*B = B satisfied\n');
	% Relation 2. BDagg*B*BDagg = BDagg
	if mod(BDagg*B*BDagg,2) == mod(BDagg,2)
	fprintf(1,`\nRelation BDagg*B*BDagg = BDagg satisfied\n');
	% Relation 3. (B*BDagg)' = B*BDagg
	if mod((B*BDagg)',2) == mod(B*BDagg,2)
	fprintf(1,`\nRelation (B*BDagg)'' = B*BDagg satisfied\n');
	end
	% Relation 4. (BDagg*B)' = BDagg*B
	if mod((BDagg*B)',2) == mod(BDagg*B,2)
	isInvFound = 1;
	fprintf(1,`\nRelation (BDagg*B)'' = BDagg*B satisfied\n');
	end
	end
	end
	else
	fprintf(1,`\nError in Inverse calculation of matrix\n');
	end
	\end{verbatim}
	
	\color{blue} \begin{verbatim}
	Original Matrix :
	B =
	00010
	10000
	01010
	00110
	
	
	Generalized Inverse Matrix :
	BDagg =
	0100
	1010
	1001
	1000
	0000
	
	Relation B*BDagg*B = B satisfied
	
	Relation BDagg*B*BDagg = BDagg satisfied
	
	Relation (B*BDagg)' = B*BDagg satisfied
	
	Relation (BDagg*B)' = BDagg*B satisfied
	\end{verbatim} \color{black}
	
	\subsection{MATLAB code for execution of ISD attack on small parameters of McEliece PKC} 
	
	\begin{verbatim}
	
	tstart = cputime;
	fid = fopen(`Example1.txt');
	code = fscanf(fid,`%d',[1,2]);
	G  = fscanf(fid,`%d',[code(1,1),code(1,2)]);
	e = fscanf(fid,`%d',[1,code(1,2)]);
	fclose(fid);
	k = code(1,1);
	n = code(1,2);
	
	fprintf(1,`Information set decoding attack for McEliece PKC.\n');
	fprintf(1,`k = %d\n',k);
	fprintf(1,`n = %d\n',n);
	
	m = randi([0,1],[1,k]);
	m
	G
	e
	
	c = mod(mod(m*G,2) + e,2);
	c
	fid = fopen(`ans1.txt',`w');
	fprintf(fid,`Information set decoding started:\n');
	nck = nchoosek(n,k);
	nckMatrix = nchoosek((1:n),k);
	for i=1:nck
	fprintf(fid,`---------------------------------------');
	fprintf(fid,`\nIteration: %d \n',i);
	fprintf(fid,`I%d = {',i);
	fprintf(fid,`%d, ',nckMatrix(i,:));
	fprintf(fid,`}\n');
	
	Gdelta = zeros(k,k);
	
	for j = 1:k
	Gdelta(:,j) = G(:,nckMatrix(i,j));
	cdelta(:,j) = c(:,nckMatrix(i,j));
	end
	[GdeltaDiag,isInvFound] = genInverse(k,k,Gdelta);
	if isInvFound == 1
	mVerify = mod(cdelta*GdeltaDiag,2);
	if all(m == mVerify) == 1
	fprintf(fid,`\nInformation set decoding attack successful.\n');
	%mVerify
	%GdeltaDiag
	%break;
	else
	fprintf(fid,`\nInformation set decoding attack unsuccessful.\n');
	end
	else
	fprintf(fid,`\nWrong Information set selected.\n');
	end
	end
	fprintf(fid,`---------------------------------------\n');
	tend = cputime - tstart;
	fprintf(fid,`Time taken in complete execution: %g\n',tend);
	fclose(fid);
	\end{verbatim}
	
	\color{blue} \begin{verbatim}Information set decoding attack for McEliece PKC.
	k = 8
	n = 16
	
	m =
	00000100
	
	G =	
	1010011010000000
	0111010101000000
	1111000000000001
	0111101000010000
	1011110000001000
	1010110100000010
	1001111000000100
	0110001100100000
	
	e =
	1010000000000000
	
	c =
	0000110100000010
	
	Output of `ans.txt':
	
	Information set decoding started:
	---------------------------------------------
	Iteration: 1 
	I1 = {1, 2, 3, 4, 5, 6, 7, 8, }
	
	Information set decoding attack unsuccessful.
	---------------------------------------------
	
	\end{verbatim}
	
	\hspace{4cm}\vdots
	\begin{verbatim}
	
	Iteration: 4 
	I4 = {1, 2, 3, 4, 5, 6, 7, 11, }
	
	Wrong Information set selected.
	---------------------------------------------
	\end{verbatim}
	
	\hspace{4cm}\vdots
	\begin{verbatim}
	Iteration: 8159 
	I8159 = {2, 4, 5, 6, 7, 8, 10, 11, }
	
	Information set decoding attack successful.
	---------------------------------------------
	
	\end{verbatim}
	
	\hspace{4cm}\vdots
	\begin{verbatim}
	---------------------------------------------
	Time taken in complete execution: 2.10938
	\end{verbatim} \color{black}

\end{appendices}

\end{document}